\documentclass[journal=jctcce,manuscript=article,chaptertitle=true]{achemso}

\usepackage[english]{babel}
\addto\captionsenglish{

}

\usepackage[sc]{mathpazo}        
\usepackage{amsmath}
\usepackage{mathrsfs}
\usepackage{amssymb}

\usepackage{adjustbox}

\usepackage{xspace}
\usepackage{booktabs}            
\usepackage[table,usenames,dvipsnames]{xcolor} 
\usepackage{amsmath}
\usepackage{amssymb}             
\usepackage{multirow}            
\usepackage{datetime2}           
\usepackage{circledsteps}        
\usepackage{diagbox}             
\usepackage{caption}
\usepackage{subcaption}
\usepackage[numbers,super,sort&compress]{natbib}
\usepackage{graphicx}            
\usepackage[textsize=footnotesize]{todonotes}
\usepackage{url}
\definecolor{LinkCol}{cmyk}{1.00 0.60 0.00 0.00} 

\definecolor{green}{rgb}{0.0, 0.8, 0.0}

\usepackage[%
colorlinks=true,
linkcolor=LinkCol,
urlcolor=LinkCol,
citecolor=LinkCol,
pdftitle={Constant pH Simulation with FMM Electrostatics in GROMACS}  
]{hyperref}

\newcommand{\gromacs}     {GROMACS\xspace}

\newcommand{\lambdadyn}   {$\lambda$-dynamics\xspace}
\newcommand{\pKa}         {\mbox{p$K_a$}\xspace}
\newcommand{\pH}          {\mbox{pH}\xspace}  
\newcommand{\back}        {environment\xspace}
\newcommand{\ffCharmmUsed}{CHARMM36m\xspace}
\newcommand{\ffAmberUsed} {Amber99sb*-ILDN\xspace}

\newcommand{\Ne}          {N^{(E)}}
\newcommand{\Ns}          {N^{(\sigma)}}
\newcommand{\Nsp}         {N^{(\sigma')}}
\newcommand{\Ss}          {S^{(\sigma)}}
\newcommand{\Ssr}         {S^{(\sigma,\rho)}}
\newcommand{\qs}          {\tilde{q}^{(\sigma)}}
\newcommand{\qbs}         {\tilde{\textbf{q}}^{(\sigma)}}
\newcommand{\qsp}         {\tilde{q}^{(\sigma')}}
\newcommand{\qsr}         {q^{(\sigma,\rho)}}
\newcommand{\qbsr}        {\textbf{q}^{(\sigma,\rho)}}
\newcommand{\Lsr}         {\tilde{\lambda}^{(\sigma,\rho)}}
\newcommand{\Csr}         {\mathcal{C}^{(\sigma,\rho)}}
\newcommand{\Qbsr}        {\textbf{Q}^{(\sigma,\rho)}}
\newcommand{\Qsr}         {Q^{(\sigma,\rho)}}
\newcommand{\Qhbsr}       {\hat{\textbf{Q}}^{(\sigma,\rho)}}
\newcommand{\Qhsr}        {\hat{Q}^{(\sigma,\rho)}}
\newcommand{\osr}         {\omega^{(\sigma,\rho)}}

\newcommand{\Vsr}         {\mathcal{V}^{(\sigma,\rho)}}
\newcommand{\Ltsr}        {\mathcal{L}^{(\sigma,\rho)}}
\newcommand{\fmmhi}       {MAHI\xspace}

\def\pics{./pictures/}          

\newcommand{\MPInat}{Theoretical and Computational Biophysics, 
Max Planck Institute for Multidisciplinary Sciences,
Am Fassberg 11,
37077 G\"ottingen,
Germany}

\author{Bartosz Kohnke}
\affiliation{\MPInat}

\author{Eliane Briand}
\affiliation{\MPInat}

\author{Carsten Kutzner}
\affiliation{\MPInat}

\author{Helmut Grubm\"uller}
\affiliation{\MPInat}
\email{hgrubmu@mpinat.mpg.de}

\newcommand{\manuscriptTitleBartosz}{Constant pH Simulation with FMM Electrostatics in GROMACS. \\
(B) GPU Accelerated Hamiltonian Interpolation }

\title{\manuscriptTitleBartosz}

\begin{document}

\begin{abstract}

\noindent%
The structural dynamics of biological macromolecules, 
such as proteins, DNA/RNA, or their complexes,
are strongly influenced by protonation changes of their typically many titratable groups, 
which explains their pH sensitivity.
Conversely, conformational and environmental changes in the biomolecule affect the protonation state of these groups.
With a few exceptions, 
conventional force field-based molecular dynamics (MD) simulations do not account for these effects, 
nor do they allow for coupling to a pH buffer.

The \lambdadyn method implements this coupling and thus allows for MD simulations at constant pH. 
It uses separate Hamiltonians for the protonated and deprotonated states of each titratable group, with a dynamic $\lambda$ variable that continuously interpolates between them. 
However, rigorous implementations of Hamiltonian Interpolation (HI) \lambdadyn are prohibitively slow for typical numbers of sites when used with Particle Mesh Ewald (PME). 
To circumvent this problem, 
it has recently been proposed to interpolate the charges instead of the Hamiltonians (QI).

Here, in the second of two companion papers, 
we propose a rigorous yet efficient Multipole-Accelerated Hamiltonian Interpolation (\fmmhi) method to perform \lambdadyn in \gromacs.
Starting from a charge-scaled Hamiltonian,
precomputed with the Fast Multipole Method (FMM), 
the correct HI forces are calculated with negligible computational overhead. 
However, other electrostatic solvers, such as PME, can also be used for the precomputation.
We compare Hamiltonian interpolation with charge interpolation and show that HI leads to more frequent transitions between protonation states, resulting in better sampling and accuracy. 
Our accuracy and performance benchmarks show that introducing, e.g., 512 titratable sites to a one million atom MD system increases runtime by less than 20\% compared to a regular FMM-based simulation. 
We have integrated the scheme into our GPU-accelerated FMM code for the simulation software \gromacs, 
allowing easy and effortless transitions from standard force field simulations to constant pH simulations.

\end{abstract}


\section{Introduction}
\label{sec:introduction}

The pH of a solution is of vital importance to biomolecules, 
as evidenced by its tight regulation in the cellular environment.
Even small deviations of 0.6 pH points from physiological values can be incompatible with life,\cite{mochizuki2021acidemia, osman2001response}
as pH controls the structural integrity of proteins\cite{dill1991denatured, talley2010ph} 
and affects important catalytic processes.\cite{kishore2012thermal, cook1981use, talley2010ph}
For a more accurate description of biomolecules by molecular dynamics (MD) simulations, proper control of pH is therefore vital. 
Analogous to controlling temperature $T$ and pressure $P$ by a thermostat and a barostat,\cite{bussi2007canonical,andersen1980molecular,ParrinelloRahman1980} respectively, 
controlling the pH would allow a dynamically changing protonation state
while---ideally---producing the same average protonation and fluctuations as under experimental conditions.

Unfortunately, with a few exceptions,\cite{huang2016all, aho2022constph} 
a computationally simple yet accurate "acidostat" for the protonation chemical potential $\mu_\mathrm{H^{+}}$
is not a common feature of MD simulation packages.
Although over the past years a number of such techniques have been proposed, 
among those discrete switching of the protonation state based on intermittent Monte Carlo moves,
\cite{Buergi2002,Baptista2002,Walczak2002,Mongan2004,Dlugosz2004,Meng2010} 
continuous switching,\cite{mertz1994molecular,baptista1997,borjesson2001,Radak2017}
and various flavors of \lambdadyn,\cite{Kong1996,Wallace2011,wallace2012charge,goh2014constant,
huang2016all,lee2004constant,khandogin2005constant,Donnini2011,Harris2022,Shen2022,aho2022constph} 
all collectively referred as constant pH MD.
These techniques often require extensive enhancements of the underlying simulation code,\cite{Shen2022} 
typically at a significant cost in computational speed and increased simulation protocol complexity.\cite{Donnini2011,dobrev2017}
Among these techniques, \lambdadyn has
emerged as the preferred approach for explicit solvent constant pH MD simulations.

Similarly to free energy perturbation (FEP)\cite{Zwanzig1954FEP} or thermodynamic integration (TI),\cite{vanGunsteren1993TI} \lambdadyn describes a system 
of interest using sub-Hamiltonians for different protonation states.
For example, in the simplest case of a single titratable molecule, 
two sub-Hamiltonians are used to represent the protonated state ($\mathcal{H}_0$) 
and the deprotonated state ($\mathcal{H}_1$).
Their combination yields the full Hamiltonian
\begin{equation}
    \label{eqn:ham_lambda}
    \mathcal{H} = (1-\lambda)\mathcal{H}_0 + \lambda\mathcal{H}_1
\end{equation}
using a continuous variable $\lambda$ to linearly interpolate between both possible end states.
This approach is referred to as \emph{Hamiltonian interpolation} (HI).
Unlike both TI and FEP, where $\lambda$ is a control parameter, 
\lambdadyn associates $\lambda$ with a mass $m$ and a velocity $\dot{\lambda}$,
making $\lambda$ a "pseudo-particle" whose time evolution is governed by an extended Hamiltonian
\begin{equation}
    \mathcal{H} = (1-\lambda)\mathcal{H}_0(\mathbf{x}) + \lambda\mathcal{H}_1(\mathbf{x}) + \frac{m}{2}\dot{\lambda}^2 + V(\lambda)
    \label{eqn:ham_lambda_with_kinetic}
\end{equation}
on par with the Cartesian coordinates $\mathbf{x}$ of all "real" particles of the system.
The term $V(\lambda)$ is essential to achieve a sufficiently accurate description and control of the protonation thermodynamics and kinetics, 
as explained in detail in our companion publication.\cite{Briand2024}
Here we will focus on aspects specific to electrostatics, 
in particular on the efficient calculation of the pseudo-force on the $\lambda$ particle
\begin{equation}
\label{eqn:rek_ham_lambda_simple}
    \frac{\partial \mathcal{H}}{\partial \lambda} = \mathcal{H}_1(\mathbf{x}) - \mathcal{H}_0(\mathbf{x}), 
\end{equation}
which is required to calculate \lambdadyn trajectories.

The calculation of long-range electrostatic forces is a notorious challenge and efficiency bottleneck 
in modern MD simulations.\cite{jung2019scaling,ohno2014petascale,Pall:2015}
These forces decrease as $1/{r^2}$ and, due to their long-range nature, 
must be computed across the entire simulation box, 
leading to an $\mathcal{O}({N^2})$ calculation scheme for $N$ particles,
which would severely limit simulation system size without more efficient approximation methods. 
The \textit{de facto} standard electrostatic solver for MD simulations,
Particle-Mesh Ewald (PME),\cite{Essmann:1995vj} relies on Fast Fourier Transforms (FFT) 
to calculate the long-range part of the electrostatic interactions of periodic systems using discretized grid-based charges.

For rigorous HI with PME, 
the evaluation of ${\partial \mathcal{H}}/{\partial \lambda}$ (eq~\ref{eqn:rek_ham_lambda_simple}) requires separate computations for the two sub-Hamiltonians $\mathcal{H}_1$ and $\mathcal{H}_0$, 
where each sub-Hamiltonian requires a separate grid and thus a separate FFT,
which is the most communication-intensive and therefore performance-limiting part of the parallel PME algorithm.

For systems with many titratable sites, PME-based HI results in a computational effort scaling linearly with the number of Hamiltonians, 
which would render it impractical for constant pH simulations with larger number of titratable groups. 
To overcome this problem, 
alternative methods such as \emph{charge interpolation} (QI) (aka charge scaling) have been proposed,\cite{lee2004constant,huang2016all,aho2022constph}
where, instead of the Hamiltonians, the partial charges are interpolated,
which allows for efficient computation of the force on the $\lambda$ particle while still using PME.
HI and QI generally produce different forces on $\lambda$ particles, 
but the implications of these differences remain poorly understood.
Although both methods have been widely used, 
further investigation has been limited by the high computational cost of HI, 
mainly due to a large overhead of additional long-range electrostatics calculations.


Despite the prevalence of FFT-based methods for long-range electrostatics in MD,  
alternatives such as the Fast Multipole Method\cite{greengard1987} (FMM) exist, 
which scales asymptotically linearly with respect to the number of particles.
The method approximates the potential and the forces with a hierarchical scheme of multipole-multipole interactions.
Due to the hierarchical decomposition of the simulation volume, 
good parallel scalability is achieved,\cite{Yokota_Turkiyyah_Keyes_2014} too, 
as the communication effort scales as $\mathcal{O}(\log P)$ with the number of computational nodes $P$ in contrast to PME, 
where $\mathcal{O}(P^2)$ requirement limits parallel scaling.\cite{Pall:2015}
In addition, the spatial decomposition of the computational domain and the ability to separate the periodic and non-periodic parts of the calculation open up new possibilities for MD simulations,
such as sparse systems like aerosols or droplets,\cite{Marklund:2024} 
systems with open boundaries,\cite{kohnke2020}
and, most importantly for this work,
Hamiltonian interpolation for \lambdadyn. 

Here we develop and assess an efficient implementation of HI-based \lambdadyn
for systems with a large number of titratable sites.
We introduce a scheme that allows an efficient calculation of larger numbers of ${\partial \mathcal{H}}/{\partial \lambda}$ values.
As a result, our implementation requires almost no additional computational effort even for large numbers of protonatable sites and, 
hence, $\lambda$ particles.
While constant pH simulations are a natural application for our method, 
FMM-based HI has the potential to go beyond this by exploiting the flexibility of both FMM and HI.
This combination allows to tackle more complex scenarios with multiple Hamiltonians that differ not only in a few charges but also, e.g., in the number of atoms.
We have integrated this scheme into our GPU-accelerated FMM code\cite{kohnke2020, bkohnke2021} for the simulation software \gromacs.\cite{Abraham2015,Pall:2015} 
As a result, as described and tested in detail in our companion publication,\cite{Briand2024} HI for constant pH MD of large protein simulation systems and long time scales 
has the potential to be used in a straightforward way,
with the simulation setup effort similar to established fixed-protonation simulations and at small runtime overhead.
To demonstrate the practical consequences of choosing HI over QI, 
we have identified some of the differences between them.

\section{Theory}

The FMM implementation used for the present work, as well as its optimizations 
and accuracy/performance evaluation have been reported previously.\cite{bkohnke2021,kohnke2020}
After a brief summary of the FMM, here we describe the FMM extensions relevant for \lambdadyn.

\subsection{Fast Multipole Method}

\begin{figure}[tbp]
\centering
\includegraphics[width=0.7\textwidth]{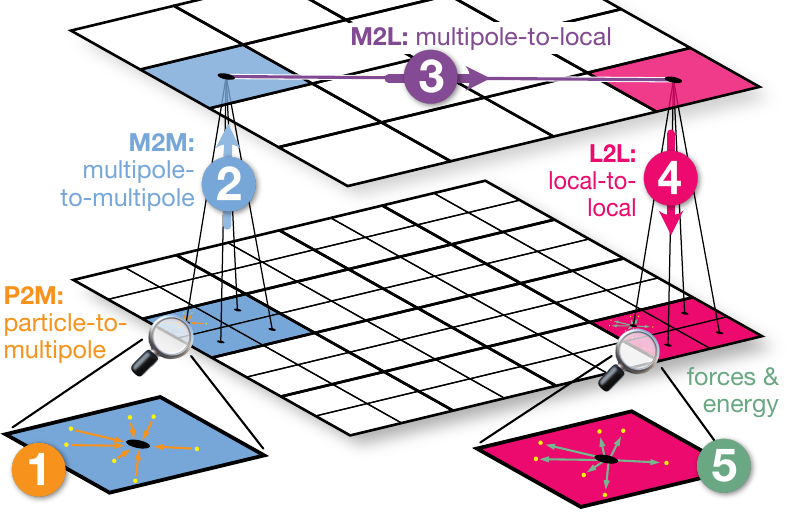}
\caption{{\bfseries FMM far field calculation.}
The five individual steps and operators involved in the far field calculation, shown for the lowest two levels of the octree:
\Circled{1} P2M: At the lowest level, the individual charges (yellow dots) are combined into a multipole representation.
\Circled{2} M2M: The multipoles of the higher levels are derived  from those of the lower levels (blue).
\Circled{3} M2L: The multipoles (blue) are transformed into local moments (pink) at each level of the tree.
\Circled{4} L2L: The local moments are propagated down the tree to the deepest level.
\Circled{5}: The local moments are used to calculate the far field contribution to the forces on the particles.
}
\label{fig:fmm_scheme}
\end{figure}

FMM approximates electrostatic interactions between $N$ particles by grouping them into a near field and a far field based on their mutual distances (see Figure~\ref{fig:fmm_scheme}). 
In the near field, particle-particle interactions (in short P2P) are directly evaluated via the Coulomb sum, 
whereas the far field interactions are approximated multipole-multipole interactions, 
truncated at a pre-specified multipole order $p$.

To group interactions into near and far field, 
the cubic simulation box is hierarchically divided into eight equally sized sub-boxes, resulting in an octree. At depth $d$, all particle-particle interactions between adjacent boxes are considered to be near field, 
whereas interactions between distant boxes are assigned to the far field.
The depth $d$ of the octree is selected based on the system particle count $N$, 
ensuring a balance between near and far field computational effort to optimize performance. 

The current version of our FMM-based constant pH implementation is limited to cubic simulation systems. 
Future updates will allow for non-cubic shapes, providing greater flexibility for a wider range of molecular systems.

\subsection{Hamiltonian Interpolation Formulation}
\label{sec:Hamiltonian_Interpolation_Formulation}
Throughout this manuscript, 
we will use the term \emph{site} for all atoms of a titratable group that change their partial charge upon protonation/deprotonation.
Accordingly, we will use the term \emph{form} to refer to a chemically distinct state of each titratable \emph{site}.
E.g., the simplest site comprises a protonated and a deprotonated form, where each form is described by a different Hamiltonian such as in eq~\ref{eqn:ham_lambda}.
Generalizing this approach, a system that contains two titratable sites can be described by recursively expanding the Hamiltonian (eq~\ref{eqn:ham_lambda}),
\begin{equation}
\label{eqn:rek_ham_lambda}
\mathcal{H} = (1-\lambda_{1})\left[(1-\lambda_{0})\mathcal{H}_{00} + \lambda_0\mathcal{H}_{01}\right] + \lambda_{1}\left[(1-\lambda_{0})\mathcal{H}_{10} + \lambda_0\mathcal{H}_{11}\right],
\end{equation}
where $\mathcal{H}_{\mathcal{X}}$, $\mathcal{X}=\{00,01,10,11\}$, describes the electrostatic interactions
\begin{equation}
\label{eqn:nbody_periodic}
\mathcal{H}_\mathcal{X} = \frac{1}{2} \sum_{\textbf{n} \in \mathbb{Z}^3} \sideset{}{'}\sum_{i,j=1}^{N} \frac{q_i^{\mathcal{X}} q_j^{\mathcal{X}}}{r_{ij} + \textbf{n}\mathrm{L}} \;,  
\end{equation} 
where both sites are protonated (00), one of two sites is protonated (01 and 10), and both sites are deprotonated (11).
Here, $\textbf{n}$ is a box shift vector, which describes the periodicity of the system,
$\mathrm{L}$ the length of the cubic simulation box, and 
$q_i^\mathcal{X}$, $q_j^\mathcal{X}$ are the partial charges of particles $i$ and $j$ according to their form and site. 
The prime at the sum symbol indicates that self-interactions, i.e.\ interactions between particles at positions $\textbf{x}_i$ and $\textbf{x}_j$, 
where $i=j$ and $\textbf{n} = 0$, are omitted. 
The recursive formulation, 
exemplified by eq~\ref{eqn:rek_ham_lambda} for two sites,
is generalized straightforwardly to systems with $M$ sites.
Note, however, that its naive implementation would require a separate evaluation of all $2^M$ sub-Hamiltonians $\mathcal{H}_\mathcal{X}$
that contain all pairwise combinations of all different forms of all sites.\cite{Donnini2011}
This approach incurs a significant computational overhead,\cite{Dongarra2021} 
and quickly becomes impractical for systems with many sites.

\begin{figure}[tbp]
\centering
\includegraphics[width=0.8\textwidth]{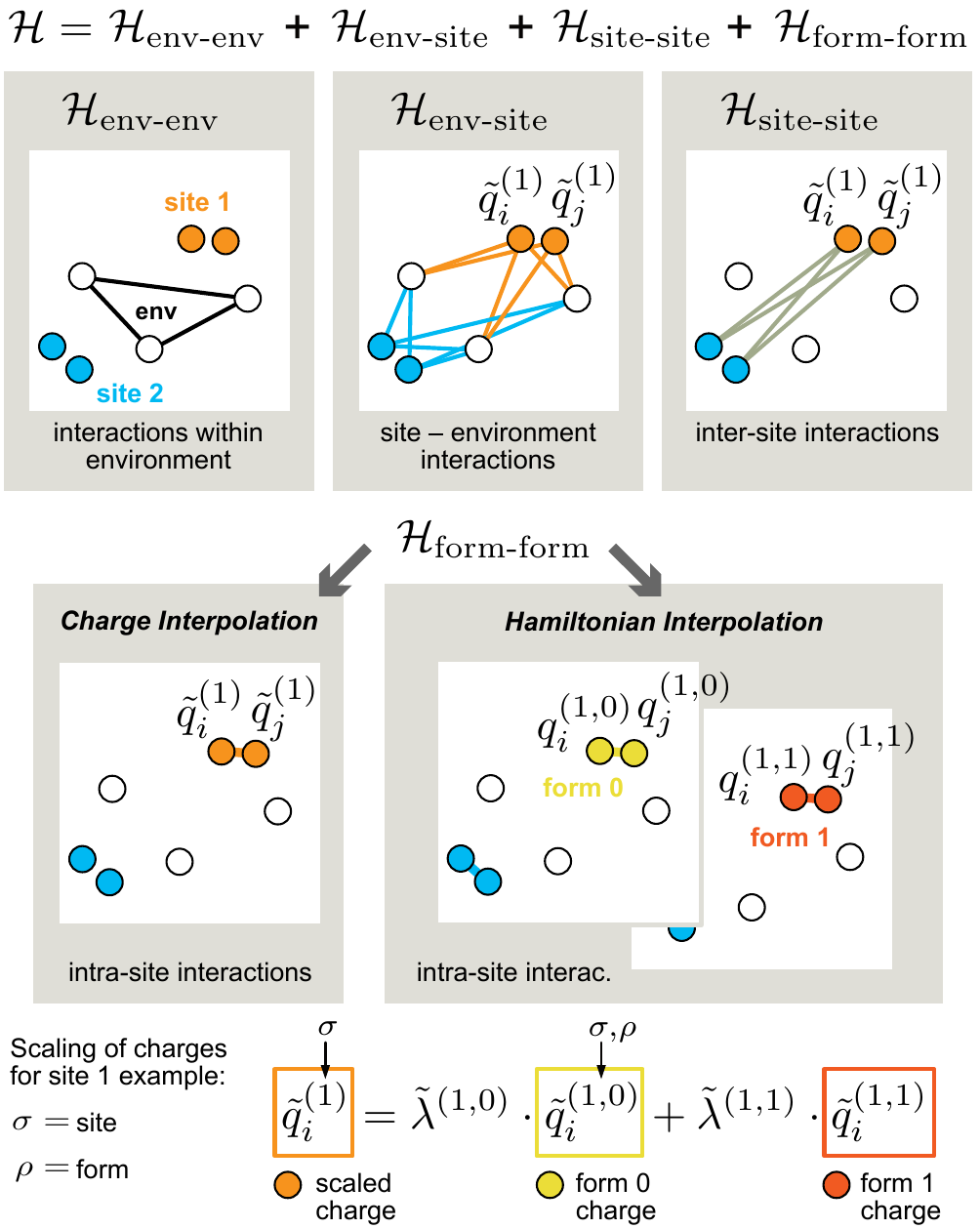}
\caption{{\bfseries Sketch of the four different types of interactions that occur in an MD system with titratable sites.}
Each gray box illustrates one term of eqs~\ref{eqn:iter_ham_lambda} and \ref{eqn:iter_ch_ham_lambda}, with particles as circles and interactions as lines.
The first three terms (top three boxes) are calculated from scaled charges ($\tilde{q}_i$, orange circles) and are identical for Hamiltonian (HI) and charge interpolation (QI).
HI differs from QI for the intra-site interactions (gray boxes in the middle),
which are calculated from scaled charges for QI (left), but from pure charges (yellow and red) for HI (right).
Scaled charges are obtained by weighing form 0 (yellow) and form 1 (red), as seen at the bottom.
}
\label{fig:interactions}
\end{figure}

To overcome this limitation, 
we switch to a mathematically equivalent formulation for describing the Hamiltonians in \lambdadyn. \cite{Hayes2024}
In {\em general formulation}, 
we consider systems with $M$ sites, 
where each site $\Ss, \sigma = 1,\dots,M$ contains $\Ns$ particles that change their partial charge upon protonation.
To allow for any number of forms $\#\Ss$ per site $S^{(\sigma)}$,
we extended the three-state model\cite{dobrev2017} to a multi-state model, 
where each site $\Ss$ contains $\Ns$ differently charged particles according to its form $\Ssr$,  $\rho = 0, \dots, \#\Ss - 1$.
For clarity of notation here and subsequently, we will omit the interactions between periodic images.
The general formulation is given by 
\begin{align}
\label{eqn:iter_ham_lambda}
\mathcal{H} &=& &\mathcal{H}_\text{env-env} + \mathcal{H}_\text{env-site} + \mathcal{H}_\text{site-site} + \mathcal{H}_\text{form-form} \\[4mm] 
&=& &\frac{1}{2}\sideset{}{'}\sum_{i,j=1}^{\Ne} \frac{{q}_i {q_j}}{r_{ij}} + 
\sum_{\sigma=1}^{M} \sum_{i=1}^{\Ns}  \sum_{j=1}^{\Ne} \frac{\qs_i q_j}{r_{ij}} + 
\sum_{\sigma=1}^{M-1} \sum_{\sigma'=\sigma+1}^{M} \sum_{i=1}^{\Ns} \sum_{j=1}^{\Nsp} \frac{ \qs_i \qsp_j}{r_{ij}} \nonumber \\ 
&+& &\frac{1}{2}\sum_{\sigma=1}^{M} \sum_{\rho=0}^{\#{\Ss} - 1}  \Lsr \sideset{}{'}\sum_{i,j=1}^{\Ns} \frac{\qsr_i \qsr_j}{r_{ij}}, \nonumber
\end{align}
where $\Ne$ is the number of non-titratable particles,
e.g. water molecules, ions or parts of the protein not affected by protonation, and
\begin{align}
\label{eqn:lambda_tilde} 
\Lsr = \mathcal{T}(\lambda_0,\dots,\lambda_{L^{(\sigma)}-1})
\end{align}
are obtained by the transformation $\mathcal{T}$,
where $L^{(\sigma)}:=\log_2(\#\Ss)$.
$\mathcal{T}$ transforms the original $\lambda$ values, as used in eq~\ref{eqn:rek_ham_lambda}, 
to $\tilde{\lambda}$ values that describe the degree to which each form of a site (site-form) is present in the system.
By construction, each $\tilde{\lambda}$ has a value between zero and one, and
\begin{align}
\label{eqn:constraint} 
\sum_{\rho = 0}^{\#\Ss - 1} \Lsr = 1, \; \sigma = 1, \dots, M.
\end{align}
The transformation is described in detail in the Appendix.
For constant pH simulations, the original $\lambda$ values reflect the progress on the protonation reaction coordinate, 
whereas the $\tilde{\lambda}$ weights describe the concentration of protonated or deprotonated species 
produced by the protonation reaction, normalized to the unit interval $[0,1]$.
Accordingly, the charges 
\begin{equation}
\qs_i = \sum_{\rho=0}^{\#\Ss-1}\Lsr \qsr_i \;, \quad \sigma=1,\dots,M\;, \quad i=1,\dots,\Ns,
\end{equation}
are $\tilde{\lambda}$ scaled charges.
In eq~\ref{eqn:iter_ham_lambda} the interactions are decomposed into four different types
(see also Figure~\ref{fig:interactions}):
\begin{itemize}

    \item[(1)] $\mathcal{H}_\text{env-env}$ contains all interactions for which none of the atoms associated 
    with the charges $q_i$ and $q_j$ are part of any site. 
    We will call the $\lambda$-independent part of the system \emph{\back}.
    These interactions do not contribute to $\partial \mathcal{H} / \partial \lambda$.

    \item[(2)] $\mathcal{H}_\text{env-site}$ contains interactions between the \back and atoms that are 
    part of a titratable site.

    \item[(3)] $\mathcal{H}_\text{site-site}$ contains interactions between atoms of different sites.

    \item[(4)] $\mathcal{H}_\text{form-form}$ contains interactions between particles 
    that belong to the same form of a titratable site. 

\end{itemize}


\subsection{Multipole-Accelerated Hamiltonian Interpolation (\fmmhi)}
\label{sec:interaction_corrections}
MD simulations of solvated biomolecules typically employ periodic boundary conditions to avoid boundary artifacts.
In addition, 
Ewald methods naturally yield intrinsic system energies, 
which are characterized by tin-foil boundary conditions at infinity.\cite{Kudin2004}
To ensure consistency with Ewald methods and to rigorously describe the specifics of \fmmhi, 
the Hamiltonian is split into three parts,
\begin{equation}
\label{eqn:approx_nbody_periodic}
\mathcal{H} \approx \frac{1}{2} \sum_{i,j=1}^{N} \sum_{\textbf{n} < 2}\frac{q_i q_j}{r_{ij} + \textbf{n}\mathrm{L}} +  \mathcal{L}(\omega) - \mathcal{D}(\omega_1).
\end{equation}
These parts are
(i) the box-box interactions of the near field and the far field, 
(ii) the periodic lattice contribution $\mathcal{L}(\omega)$, and 
(iii) the dipole compensation $\mathcal{D}(\omega_1$).
The lattice far field operator $\mathcal{L}(\omega)$ approximates the periodic interactions between the simulation box and its infinite number of copies.
The dipole compensation is a function of the dipole $\omega_1$ of the simulation box and it
ensures that the tin-foil boundary conditions at infinity are met, 
so that the energies match those obtained by Ewald methods.

To describe \fmmhi, 
we consider a system with $M$ sites $\Ss$, where $\sigma = 1,\dots,M$.
Each site $\Ss$ contains $\Ns$ particles, which can vary in their partial charge according to their form $\Ssr$, $\rho = 0,\dots,\#\Ss-1$.
First, the charge-scaled Hamiltonian
\begin{equation}
\label{eqn:scaled_electrostatics}
\tilde{\mathcal{H}} = \frac{1}{2} \sum_{\textbf{n} \in \mathbb{Z}^3} \sideset{}{'}\sum_{i,j=1}^{N} \frac{\tilde{q}_i \tilde{q}_j}{r_{ij} + \textbf{n}\mathrm{L}} \; 
\end{equation}
for $N = \Ne + N^{(1)}+\dots+N^{(M)}$ particles is calculated.
Since this calculation is performed on scaled charges $\tilde{q}_i$, 
the grouping into $M$ sites does not affect this stage of the calculation.
Consequently, 
the calculation is as efficient as in a fixed-protonation simulation with the same number of particles $N$.
Note that the precomputation of $N^{(1)}+\dots+N^{(M)}$ scaled charges $\tilde{q}_i$ is negligible in performance,
and for all \back particles $\tilde{q}_i = q_i$ holds.

To describe the next step of \fmmhi, we consider the difference between $\tilde{\mathcal{H}}$ and $\mathcal{H}$.
To this end, the charge-scaled Hamiltonian is rewritten to emphasize the grouping into $M$ sites according to the general formulation 
(eq~\ref{eqn:iter_ham_lambda}), which yields
\begin{equation}
\label{eqn:iter_ch_ham_lambda}
\tilde{\mathcal{H}} = \frac{1}{2}\sideset{}{'}\sum_{i,j=1}^{\Ne} \frac{{q}_i {q_j}}{r_{ij}} + 
\sum_{\sigma=1}^{M} \sum_{i=1}^{\Ns}  \sum_{j=1}^{\Ne} \frac{\qs_i {q}_j}{r_{ij}} + 
\sum_{\sigma=1}^{M-1} \sum_{\sigma'=\sigma+1}^{M} \sum_{i=1}^{\Ns} \sum_{j=1}^{\Nsp} \frac{ \qs_i \qsp_j}{r_{ij}} + 
\frac{1}{2}\sum_{\sigma=1}^{M} \sideset{}{'}\sum_{i,j=1}^{\Ns} \frac{\qs_i \qs_j}{r_{ij}}.
\end{equation}
This differs from the general formulation
only in the last term (see also Figure~\ref{fig:interactions}), 
which describes intra-form interactions, 
i.e.,\ interactions between particles of the same form within a site.
Hence, to retrieve the Hamiltonian $\mathcal{H}$ (eq~\ref{eqn:iter_ham_lambda}), only the intra-form interactions of the charge-scaled Hamiltonian $\tilde{\mathcal{H}}$ need to be modified.
For this purpose, the corrections
\begin{equation}
\label{eqn:correction_scheme}
\Csr = \Csr_{\text{P2P}} + \Csr_{\mathcal{L}} + \Csr_{\mathcal{D}}, \quad \sigma=1,\dots,M\;, \quad \rho=0,\dots,\#\Ss - 1
\end{equation}
are applied, which, according to eq~\ref{eqn:approx_nbody_periodic}, 
separately target 
(i) the box-box interactions (near field and far field), 
(ii) the lattice interactions, 
and (iii) the dipole compensation.
In the following each of the three different correction steps will be described in detail.


For compact notation, the correction charges for each form $\Ssr$ of $M$ sites $S^{\sigma}$ are abbreviated by
\begin{equation}
\label{eqn:charge_correction}
\Qbsr := \left(\rule{0cm}{10px}\Qsr_i = \qs_i - \frac{1}{2}{\qsr_i}, \;\;\; i = 1,\dots,\Ns\right)\rule{0cm}{10px}, 
\end{equation} 
and the dipole compensation correction charges by
\begin{equation}
\label{eqn:charge_correction_dipole}
\Qhbsr := \left(\rule{0cm}{12px}\Qhsr = \qs_i - \qsr_i, \;\;\; i = 1,\dots,\Ns\right)\rule{0cm}{12px}.
\end{equation}
The multipole expanded at the center of the simulation box with charges 
\begin{equation}
    \qbsr := (\qsr_i, \;\;\; i = 1,\dots,\Ns)
\end{equation}
is defined as
\begin{equation}
\osr := \omega(\qbsr) = \sum^{\Ns}_{i=1} \mathcal{M}(\qsr_i), 
\end{equation}
where $\mathcal{M}$ is an operator calculating the multipole expansion of the simulation box.\cite{bkohnke2021}
The local expansion of the simulation cell is obtained via the lattice operator
\begin{equation}
\Ltsr := \mathcal{L}(\omega(\qbsr)).
\end{equation}

\subsubsection{Box-Box Interactions Correction}
The box-box correction terms are calculated for each site-form as
\begin{equation}
\Csr_{\text{P2P}} = \sum_{i=1}^{\Ns} \Vsr_i \qsr_i,
\end{equation}
where 
\begin{equation}
\Vsr_i = \sum_{j=1}^{\Ns} \frac{\Qsr_j}{r_{ij}}, \;\;\; i = 1,\dots,\Ns
\end{equation}
is a correction potential evaluated between atoms within a site (i.e., intra-site). 
Since the number of particles $\Ns$ per site is typically small, 
the calculation of the correction potential $\Vsr$ has only a negligible computational overhead and can therefore be  
computed directly by evaluating the particle-particle interactions. 
Note that the interactions are calculated independently for each site-form, 
which leads to a straightforward parallelization of this correction part. 

\subsubsection{Lattice Correction}
\label{sec:Lattice_Correction}
The correction of the lattice part is calculated as 
\begin{equation}
\label{eqn:C_B}
 \Csr_{\mathcal{L}} = \osr\mathcal{L}(\omega(\Qbsr)).
\end{equation} 
The computational costs of $\osr$ and $\omega(\Qbsr)$ operations are of the order of $\mathcal{O}(\Ns)$, 
so they are negligible in runtime, since $\Ns$ is typically small (5--15 particles). 
\begin{figure}[tbp]
    \centering
    \includegraphics[width=0.65\textwidth]{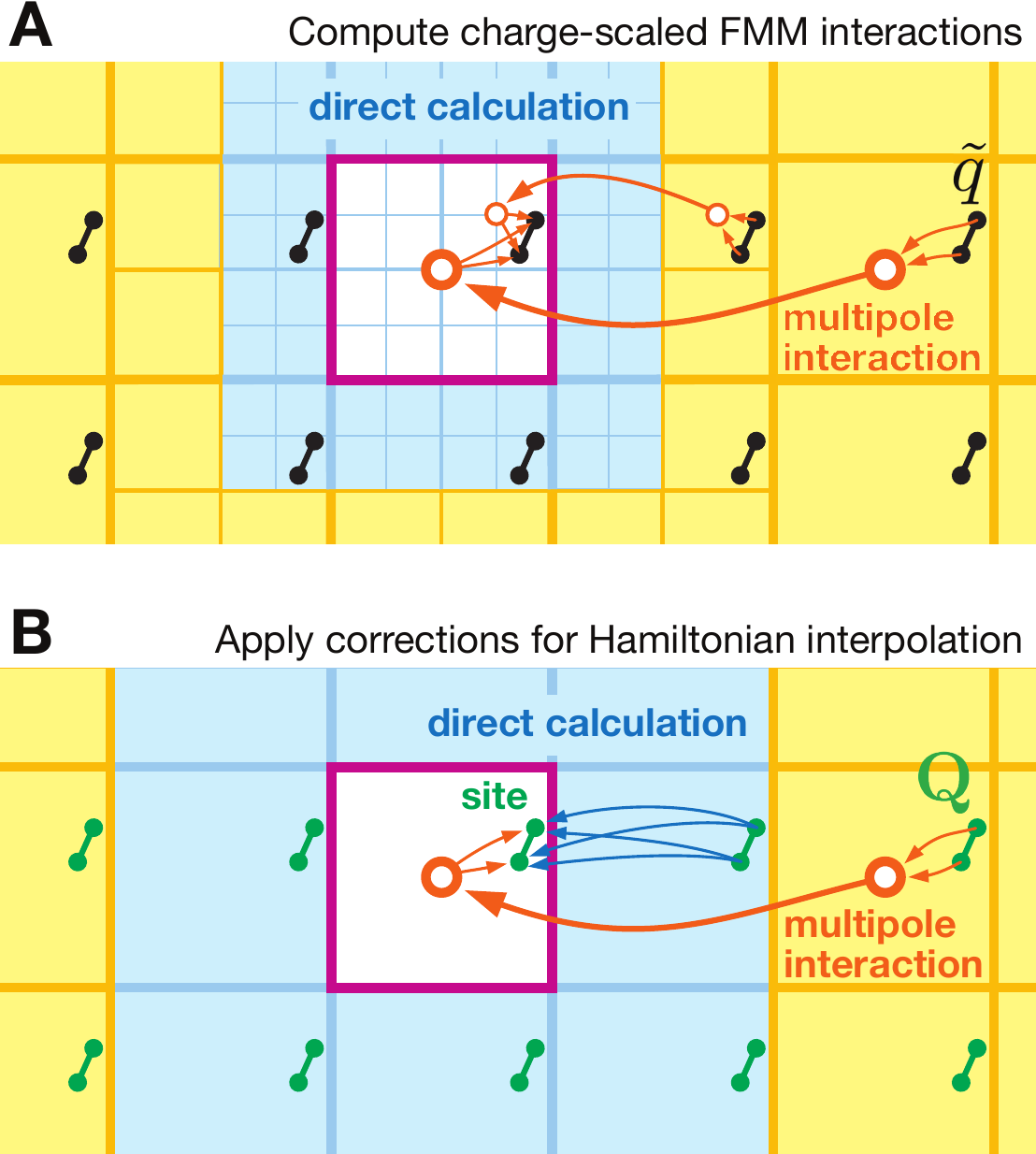}
    \caption{{\bfseries Starting from a charge-scaled Hamiltonian, 
    the \fmmhi scheme calculates the correct (periodic) $\mathcal{H}_\text{form-form}$ interactions for Hamiltonian interpolation (HI).}
    The central magenta box shows the actual simulation volume containing a two-atomic site (black/green dots), while the surrounding boxes are periodic images.
    \textbf{A.} First, FMM calculates the interactions for the scaled charges $\tilde{q}$
    using multipole expansion in the yellow areas.
    \textbf{B.} Corrections are then computed so that HI
    is retrieved for a site with charges $\text{Q}_{\text{P2P}}$ and $\text{Q}_\mathcal{L}$ (green).
    Here, in contrast to a regular FMM, all corrections to interactions coming from the first layer around the central box are computed directly (blue), while corrections from distant boxes are handled by a lattice operator (yellow).}
    \label{fig:FmmCorrections}
\end{figure}
The $\mathcal{O}(p^4)$ lattice operator $\mathcal{L}(\omega(\Qbsr))$, 
which is the most computationally expensive part of the far field evaluation, 
is also negligible in runtime even though it must be computed for each site-form.
This is because the total number of site-forms in a typical constant pH MD system, 
and therefore the number of applications of the correction lattice operator, 
is expected to be a small fraction of all far field operators used by FMM.\cite{bkohnke2021} 
Notably, similar to the box-box interactions correction $\Csr_{\text{P2P}}$,
all lattice operations for different site-forms are independent and are readily parallelized using the existing unmodified CUDA lattice operator kernels.

\subsubsection{Dipole Compensation Correction}
\label{sec:Dipole_Compensation_Correction}
The dipole compensation $\mathcal{D}(\cdot)$ (eq~\ref{eqn:approx_nbody_periodic}), 
which contributes to the total energy of a system, 
is evaluated as described elsewhere.\cite{Kudin2004}
The correction for the dipole compensation
\begin{equation}
\Csr_{\mathcal{D}} = -\frac{1}{2}\left[\rule{0cm}{10px} \omega(\Qhbsr) - \omega(\tilde{\textbf{q}})\right] \left[\rule{0cm}{12px} \mathcal{L}(\omega(\qbsr_c)) - \mathcal{L}(\omega(\tilde{\textbf{q}}))) \right]
\end{equation}
must also be applied.
To perform this operation, 
it is necessary to evaluate all multipoles $\omega(\Qhbsr)$ and $\omega(\qbsr_c)$ for each site-form. 
The evaluation of $\omega(\Qhbsr)$ depends on $\Ns$ and is therefore insignificant in runtime. 
The evaluation of $\omega(\qbsr_c)$ is performed on fictitious charges for the dipole correction ($\qbsr_c$), 
which are placed at the corners of the simulation box for the purpose of calculating the dipole compensation.\cite{Kudin2004} 
There are 50 such charges,
so this operation also does not markedly contribute to the total runtime. 
Although the correction requires an evaluation of the charge-scaled multipole $\omega(\tilde{\textbf{q}})$, 
this calculation is required only once for the whole system, so performance is not affected.
The $\mathcal{O}(p^4)$ lattice operation $\mathcal{L}(\omega(\qbsr_c))$ is performed for each site-form,
rendering the computational overhead identical to that of the lattice correction. 
Another lattice operation $\mathcal{L}(\omega(\tilde{\textbf{q}}))$ is evaluated only once for the whole system because it depends on $\omega(\tilde{\textbf{q}})$.
When the dipole compensation is engaged, 
an additional adjustment is required for the lattice term $\Csr_\mathcal{L}$, 
given by 
\begin{equation}
\Csr_{\mathcal{L}} = \Csr_\mathcal{L} + \omega(\qbsr)\mathcal{L}(\qbs).
\end{equation}
This correction leverages also the precomputed term $\mathcal{L}(\omega(\tilde{\textbf{q}}))$, 
rendering the total computational costs negligible.

\subsubsection{Execution of \fmmhi}

Figure~\ref{fig:FmmCorrections} illustrates how the corrections are applied. 
The FMM computes interactions between particles in the same box and between neighboring boxes at the deepest level $d$ of the octree directly, 
while the remaining interactions are evaluated via far field operators.
In contrast, the $\mathcal{C}_{\text{P2P}}$ part of \fmmhi calculates all interactions between particles in the central simulation box and their corresponding first periodic images as direct interactions.
This is equivalent to a FMM run at tree depth $d=0$, 
and it maximizes the performance of \fmmhi by avoiding unnecessary use of $\mathcal{O}(p^4)$ operators for typically few particles of a titratable site.
Only the more distant periodic images are corrected with the corresponding lattice operators. 

\subsubsection{Calculation of Forces on $\lambda$ Particles}

To compute the forces on the original $\lambda$ particles using the correction terms $\Csr$, 
an additional step is required to map the forces $\partial \mathcal{H} / \partial \tilde{\lambda}$ to $\partial \mathcal{H} / \partial \lambda$. 
In general, this mapping uses index tuples obtained from the transformation $\mathcal{T}$ (eq~\ref{eqn:lambda_tilde}, 
described in detail in the Appendix). The forces are transformed according to
\begin{equation}
\frac{\partial \mathcal{H}}{\partial \lambda_i} = \mathcal{K}(\tilde{\mathcal{V}}, \mathcal{C}^*, q^*) \;,\; i = 0, \dots, L^{(\sigma)} - 1, \;\; \sigma = 0, \dots, M,
\end{equation}
where $\tilde{\mathcal{V}}$ is the charge-scaled potential, 
$\mathcal{C}^*$ is a list of all correction terms, 
and $q^*$ represents the site-form particles present in the system. 
The mapping $\mathcal{K}(\tilde{\mathcal{V}}, \mathcal{C}^*, q^*)$ is described in detail in the Appendix.

\subsubsection{Complexity Evaluation of \fmmhi}
To evaluate the computational complexity of \fmmhi, consider a system of $N$ particles. 
The subdivision of the system into $M$ sites $\Ss$, where $\sigma = 1,\dots,M$, 
does not increase the total number of particles. 
Thus, the computational complexity of the electrostatics solver used for the precalculation of $\tilde{\mathcal{H}}$ does not depend on the subdivision $N = N^{(E)} + N^{(1)}, \dots, N^{(M)}$, 
and therefore remains constant with respect to the growing number of sites.
The number of applied corrections $\Csr$, however, 
depends on the number of site-forms $\mathcal{F}$.
Since both the $\Ns$-- dependent part $\Csr_{\text{P2P}}$, 
as well as the $\Ns$-- independent parts $\Csr_{\mathcal{L}}$ and $\Csr_{\mathcal{D}}$,
require only a constant amount of work independent of all other site-forms,
the overall complexity of \fmmhi is $\mathcal{O}(\mathcal{F})$.

\subsection{Comparison of Hamiltonian and Charge Interpolation}
\label{sec:HIvsQI_theory}
Before presenting the accuracy and performance benchmarks for our \fmmhi, we highlight the main differences between the Hamiltonian Interpolation (HI) implemented here and charge interpolation (QI). Both methods have been successfully used for \lambdadyn simulations.
While early work tended to focus on HI, 
recent \lambdadyn implementations have turned to QI as it can be efficiently implemented using the PME electrostatic solver.\cite{aho2022constph,Harris2022,Briand2024} 

As can be seen from comparing the QI Hamiltonian $\mathcal{\tilde{H}}$ (eq~\ref{eqn:iter_ch_ham_lambda}) with the HI Hamiltonian (eq~\ref{eqn:iter_ham_lambda}), the respective forces
$\tilde{F}_{\lambda} := -{\partial\mathcal{\tilde{H}}}/{\partial\lambda}$ and ${F}_{\lambda} := -{\partial\mathcal{{H}}}/{\partial\lambda}$ 
differ only by intra-site interactions
($\mathcal{H}_\text{form-form}$ in Figure~\ref{fig:interactions},
see Section~\ref{sec:interaction_corrections}).
Typically, these involve chemically directly bonded atoms, or atoms bonded to a common neighbor, 
and are therefore excluded from the calculation.\cite{Braun2018}
Other intra-site interactions are not excluded, however, and therefore do matter, 
such as the 1-4 interactions between the proton and the other oxygen in the carboxyl group of aspartic acid (Asp) or glutamic acid (Glu).
Also interactions between the proton and the other nitrogen in the imidazole moiety in histidine (His), 
or the backbone oxygen and nitrogen with the side chain heteroatoms, are typically not excluded and therefore, too, contribute to differences between HI and QI.
Further differences are caused by interactions between otherwise excluded atoms and their periodic images.

To explore the differences resulting from form-form interactions in more detail, 
consider a single site with two forms differing in $n$ partial charges.
Again omitting interactions between periodic images for brevity of notation, 
the correction term simplifies to
\begin{equation}
\frac{\partial\mathcal{\tilde{H}}}{\partial\lambda} = \frac{\partial\mathcal{H}}{\partial\lambda} + (\lambda - \frac{1}{2}) 
\underbrace{\sum_{i,j=1}^{n} {(q_i^{(0)} - q_i^{(1)})(q_j^{(0)} - q_j^{(1)})} \frac{1}{r_{ij}}}_{k(\mathbf{r})} \ ,
\label{eqn:DeltaFQIHI}
\end{equation}
where $r_{ij}$ is the distance between atoms $i$ and $j$ of the site.
As the charges of the end states $q_i^{(0)}$, $q_i^{(1)}$, $q_j^{(0)}$, and $q_j^{(1)}$ do not change, 
the factor $k(\mathbf{r})$ depends only on the atomic positions. Integrating over $\lambda$ yields 
%
%
\begin{equation}
\tilde{\mathcal{V}}(\lambda) - \mathcal{V}(\lambda) = \frac{k(\mathbf{r})}{2}(\lambda^2 - \lambda)
\label{eqn:Udiff}
\end{equation}
as a $\lambda$ dependent potential difference between QI and HI.
This difference is a harmonic potential centered at $\lambda = 0.5$, 
with a force constant $k(\mathbf{r})$, 
acting as an additional barrier (or well) in QI, which is absent in HI.
The barrier vanishes only at exactly $\lambda = 0$ and $\lambda = 1$. 
However, "protonated" or "deprotonated" states correspond to ensembles of states around these 
values of $\lambda$. 
Consequently, the QI and HI Hamiltonians and their corresponding free energies do differ.
Therefore, we expect HI and QI to behave differently.
In section~\ref{sec:HIvsQI} we will see that this difference between QI and HI affects the protonation/deprotonation kinetics and thus the convergence of constant pH simulations.

\section{Accuracy and Performance Assessment}
We have shown earlier\cite{kohnke2020} that the electrostatic potential and the forces 
calculated using our FMM implementation within the \gromacs suite for non-periodic boundary conditions
approach the analytical solution with increasing multipole order $p$.
For periodic boundary conditions (PBC), FMM with multipole order $p=8$ achieves the same single precision accuracy for both energies and forces as PME with standard parameters for typical MD systems.
We have also shown that for multipole order $p=50$ our FMM implementation yields the analytic solution for a periodic lattice system within double precision accuracy.

To demonstrate that also our \fmmhi yields correct forces on the $\lambda$ particles in PBC settings,
we compared $\partial\mathcal{H} / \partial\lambda_i$ with reference values 
obtained from regular FMM (or PME) electrostatics 
for individual Hamiltonians (i.e.\ for $\lambda_i = 0$ and $\lambda_i = 1$).
Both individual Hamiltonians were then used to obtain total energies for arbitrary intermediate $\lambda$ values 
to check the accuracy of our method between these end states.

Furthermore, we examined how the computational performance of \fmmhi scales with the number of titratable sites and forms.
For constant pH simulations under physiological conditions (i.e.\ pH $\approx$ 7),
most sites comprise either two (Asp, Glu) or three (His) forms. 
To also assess computational performance for future generalized applications, 
we considered up to 16 forms per site.


\subsection{Description of the Benchmark Systems}
Here we briefly describe the benchmark systems that were used to evaluate the accuracy and performance of our implementation. 

\subsubsection{Random Systems}
Random systems comprised 1,000 \back particles and 10 site particles with charges drawn from a uniform distribution between $-1$ and $+1$.
Two systems were constructed to evaluate a \textit{typical case} and a hypothetical \textit{worst-case}.
For the typical case, the site particles were positioned in close proximity to each other, 
mimicking typical biological systems where particles of the same site usually belong to a single amino acid. 
For the worst case, site particles were uniformly distributed within the entire simulation box. 
While quite unrealistic, 
the latter system provides lower bound for simulation accuracy and efficiency.
The same systems were used for tests with two and four forms.

To assess the accuracy of \fmmhi, 
the forces on the $\lambda$ particles were compared with reference forces.
The latter were obtained by computing required Hamiltonians $\mathcal{H}_\mathcal{X}$ with a separate FMM run
and using
\begin{equation}
\label{eqn:ref_two_forms}
F^\text{ref}_{\lambda} := - \frac{\partial\mathcal{H}}{\partial\lambda} = \mathcal{H}_{0} - \mathcal{H}_{1}
\end{equation}
for the system with two forms and 
\begin{align}
\label{eqn:ref_four_forms}
F^\text{ref}_{\lambda_0} := -\frac{\partial\mathcal{H}}{\partial\lambda_0} = (1 - \lambda_1)\mathcal{H}_{00} + \lambda_1(\mathcal{H}_{10} + \mathcal{H}_{01} + \mathcal{H}_{11}) - \mathcal{H}_{10}\\
F^\text{ref}_{\lambda_1} := -\frac{\partial\mathcal{H}}{\partial\lambda_1} = (1 - \lambda_0)\mathcal{H}_{00} + \lambda_0(\mathcal{H}_{10} + \mathcal{H}_{01} + \mathcal{H}_{11}) - \mathcal{H}_{01}
\end{align} 
for the system with four forms. 
Subsequently, the relative deviations 
$(F_{\lambda} - F^\text{ref}_{\lambda}) / F^\text{ref}_{\lambda}$ were determined at different multipole orders $p$ and tree depths $d$.
Note, that the reference and \fmmhi forces were calculated at the same multipole order $p$.
This exactly quantifies the \fmmhi forces deviations from those obtained with naively computed HI, 
while avoiding the force differences emerging due to different precision levels.
The $\lambda$ values were kept constant at randomly chosen $\lambda=0.345$ for the system with two forms and 
at $\lambda_0=0.345$ and $\lambda_1=0.721$ for the system with four forms. 

\subsubsection{Random Systems with Varying Number of Sites and Forms}
To quantify the scaling behavior of our \fmmhi scheme with respect to the number of sites and the number of forms per site, 
we set up various random systems with 250 to 12,725,399 particles, one to 512 sites, and two to 16 forms per site.

\subsubsection{Random System with Varying Number of Particles}
To quantify how the computational effort of \fmmhi scales with the total number of charges in a system, 
we considered a series of random position systems with particle numbers ranging from 250 to 33,554,432. 
A typical fraction of one 10-atom titratable site per 4,000 atoms was chosen in each case, 
estimated from a solvated globular lysozyme protein.\cite{ramanadham1990refinement}

\subsubsection{Benzene Ring in Water}
\label{sec:benzene}
To test the overall accuracy of the $\lambda$ forces provided by \fmmhi within \gromacs,
we considered a solvated benzene molecule, 
comprising a C$_{6}$H$_{6}$ ring and 2,161 TIP3P water molecules\cite{jorgensen1983comparison}
in a $4\times4\times4~\text{nm}^3$ box using the \ffAmberUsed force field.\cite{best2009,Lindorff2010}
The reference ${\partial\mathcal{H}}/{\partial\lambda}$ values were calculated for $\lambda$ covering the range between zero and one, 
where the benzene molecule carries its full charge at $t=0.0$~ps ($\lambda = 0$), 
while it is completely uncharged at $t=2.0$~ps ($\lambda = 1$).
All the reference values were obtained by \gromacs thermodynamic integration (TI) using PME electrostatics with 4th order B-spline interpolation, 0.12~nm grid spacing, and 1.1~nm cutoffs.     
For the FMM test runs, $p=8$ and $d=3$ were used. 
For all simulations, 
a 2~fs time step was used while constraining the bonds of the water molecules with the SETTLE algorithm,\cite{miyamoto1992settle}
and all other bonds with LINCS.\cite{Hess1997}

\begin{figure}[h]
    \centering
    \includegraphics[width=0.35\textwidth]{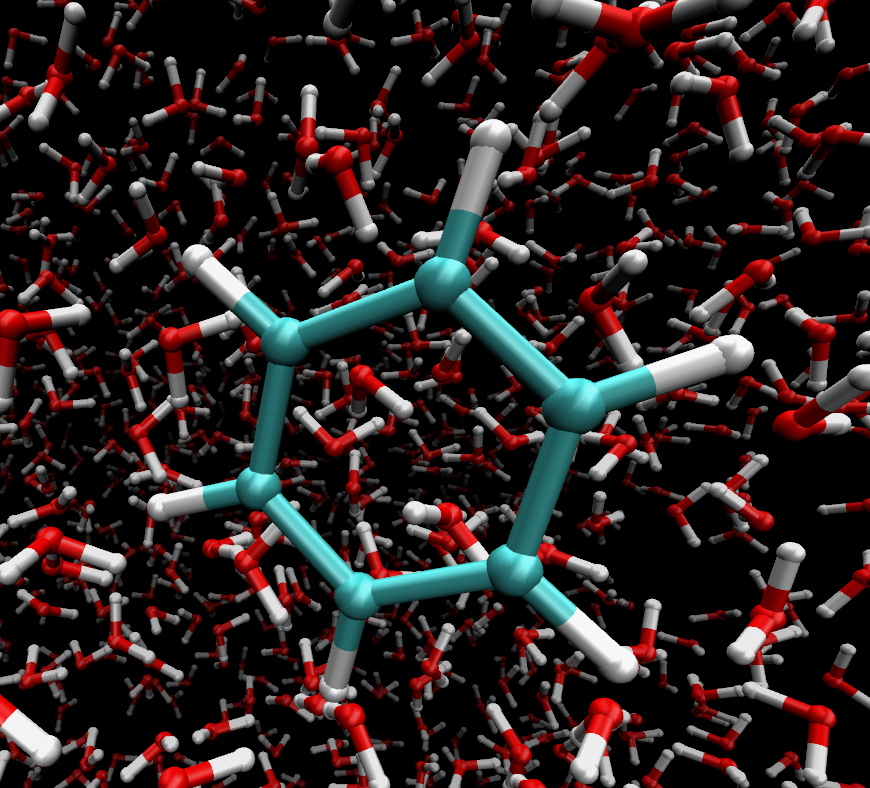}
    \caption{{\bfseries Benzene ring solvated in water.}
    The ball-and-stick drawing shows hydrogen atoms in white, carbon atoms in cyan, and oxygen atoms in red.}
    \label{fig:benzenering}
\end{figure}

\subsubsection{Constant pH Simulation Systems in \gromacs}
To assess the computational overhead of the entire constant pH \gromacs implementation relative to fixed charge FMM simulations, 
several simulation systems containing a protein with one or more titratable sites 
solvated in TIP3P\cite{huang2017} water and Na$^+$ and Cl$^-$ ions (150 mM) were considered.
These tests used the \ffCharmmUsed force field,\cite{huang2017} a 2~fs time step and a 1.2 nm van der Waals interaction cutoff.

The simplest systems contained a single solvated titratable glutamic acid (Glu) residue within cubic boxes of edge lengths 
$5~\text{nm}$, $6~\text{nm}$, $7~\text{nm}$, and $8~\text{nm}$,
comprising in a total of 12,125 20,996 33,552 and 50,682 atoms, respectively. 
To assess how the addition of titratable sites affects overall \gromacs performance, 
solvated hen egg lysozyme (PDB code 2LZT)\cite{ramanadham1990refinement} 
and staphylococcal nuclease (SNase) mutant $\Delta\text{PHS}$ (PDB code 3BDC)\cite{Castaneda2009} 
with different numbers of titratable sites were benchmarked.
Both proteins contain numerous histidine (His), aspartic (Asp) and glutamic acid (Glu) residues.
The lysozyme systems contain 1--10 protonatable residues in a 
$6.5\times6.5\times6.5~\text{nm}^3$ box,
totaling 26,761 -- 26,779 atoms.
The SNase system contains 50,749 -- 50,760 atoms
in an $8.0\times8.0\times8.0~\text{nm}^3$ box with 1--20 protonatable residues.

\subsubsection{Hamiltonian Interpolation vs. Charge Interpolation}

To characterize the differences between HI and QI for typical simulation systems, 
the relations derived in Subsection \ref{sec:HIvsQI_theory} were verified numerically.
For this purpose, the FMM code was modified to perform either HI or QI \lambdadyn,
while simultaneously reporting both ${\partial\mathcal{H}}/{\partial\lambda}$ 
and ${\partial\tilde{\mathcal{H}}}/{\partial\lambda}$.

The effect of this difference on protonation/deprotonation kinetics was assessed by counting the number of transitions between protonation states during simulations of equal lengths.
To this end, a single methyl-blocked Glu residue solvated in water was used as a test system. 
For the sake of simplicity, a two-state model without tautomery was used, and only the Glu residue was made protonatable.


First, ten replicas of the Glu system for 30~ns in \ffCharmmUsed\cite{huang2017} 
with a fixed double-well barrier height of 5 kJ/mol were simulated at pH 4.4 (the \pKa of Glu) with both QI and HI.
To ensure unbiased comparison of transition rates, 
the $V_\text{MM}(\lambda)$ potential was calibrated for a flat energy landscape at \pH = \pKa in both HI and QI as well.
Details on the $V_\text{MM}$ potential and the calibration process are in our companion publication.\cite{Briand2024}
As a typical protein system, 
SNase was simulated with the same protocol and conditions as the single residue, using 40 replicas of 60~nanoseconds each.
 
\subsection{Benchmarking Procedure}
All benchmarks were run on a compute node with an 
NVIDIA GeForce RTX 4090 GPU and 
an AMD Ryzen Threadripper 1950X 16-core processor 
with 32~GB of RAM running Scientific Linux 7.9. 
\gromacs with FMM was compiled using GCC 9.4.0 and CUDA 12.2, 
thread-MPI, and AVX2\_256 SIMD instructions, 
and with OpenMP and hwloc 2.1.0 support. 
The benchmarks with \gromacs used one thread-MPI rank and 16 OpenMP threads and were run for several thousand time steps.
Because memory allocation and load balancing typically slow down the first few hundred time steps, 
timings were collected only for the second half of each run.
All reported performances are averages over three runs.
Each of the FMM standalone benchmarks was averaged over several to several thousand runs,
depending on tree depths and particle counts.

\section{Results and Discussion}
\label{sec:results}

\subsection{Accuracy}
In order to test the accuracy of the proposed scheme,
the forces $\partial\mathcal{H} / \partial\lambda$ obtained with corrected FMM were compared with reference solutions (eq~\ref{eqn:ref_two_forms} and eq~\ref{eqn:ref_four_forms}) generated by direct (redundant) computations of the separate Hamiltonians.
Additionally, FMM-derived forces $\partial\mathcal{H} / \partial\lambda$ were compared to those from a 
standard \gromacs TI simulation using two PME calls (see Section~\ref{sec:benzene}).

\subsubsection{Accuracy of the Random System with Two Forms and Four Forms} 
\label{sec:deviation_from_reference}

\begin{figure}[tbp]
    \centering
    \begin{subfigure}[b]{0.495\textwidth}
        \centering
        \includegraphics[width=\textwidth]{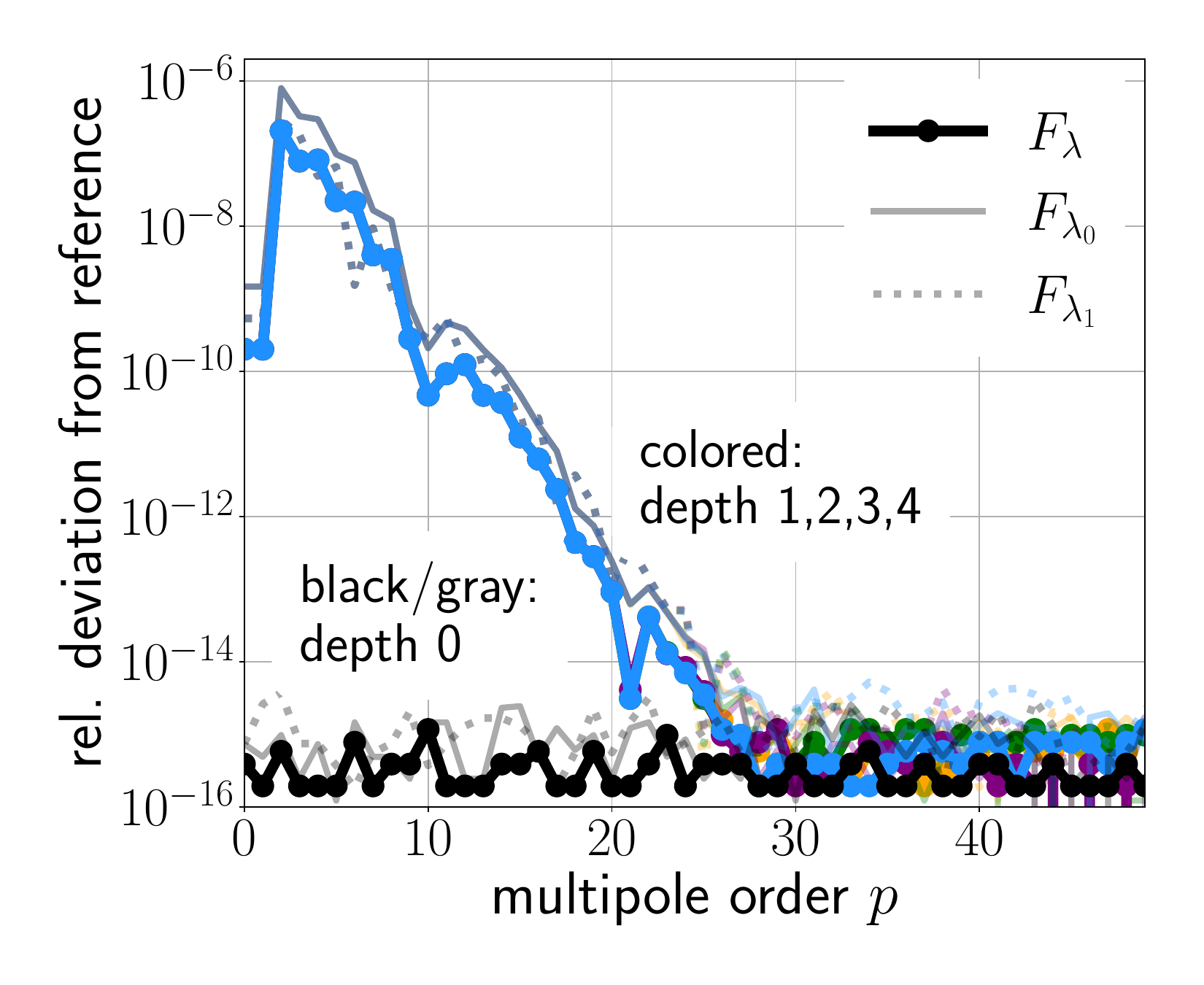}
        \caption{Double precision.}
        \label{fig:accuracy_double}
    \end{subfigure}
    \begin{subfigure}[b]{0.495\textwidth}
        \centering
        \includegraphics[width=\textwidth]{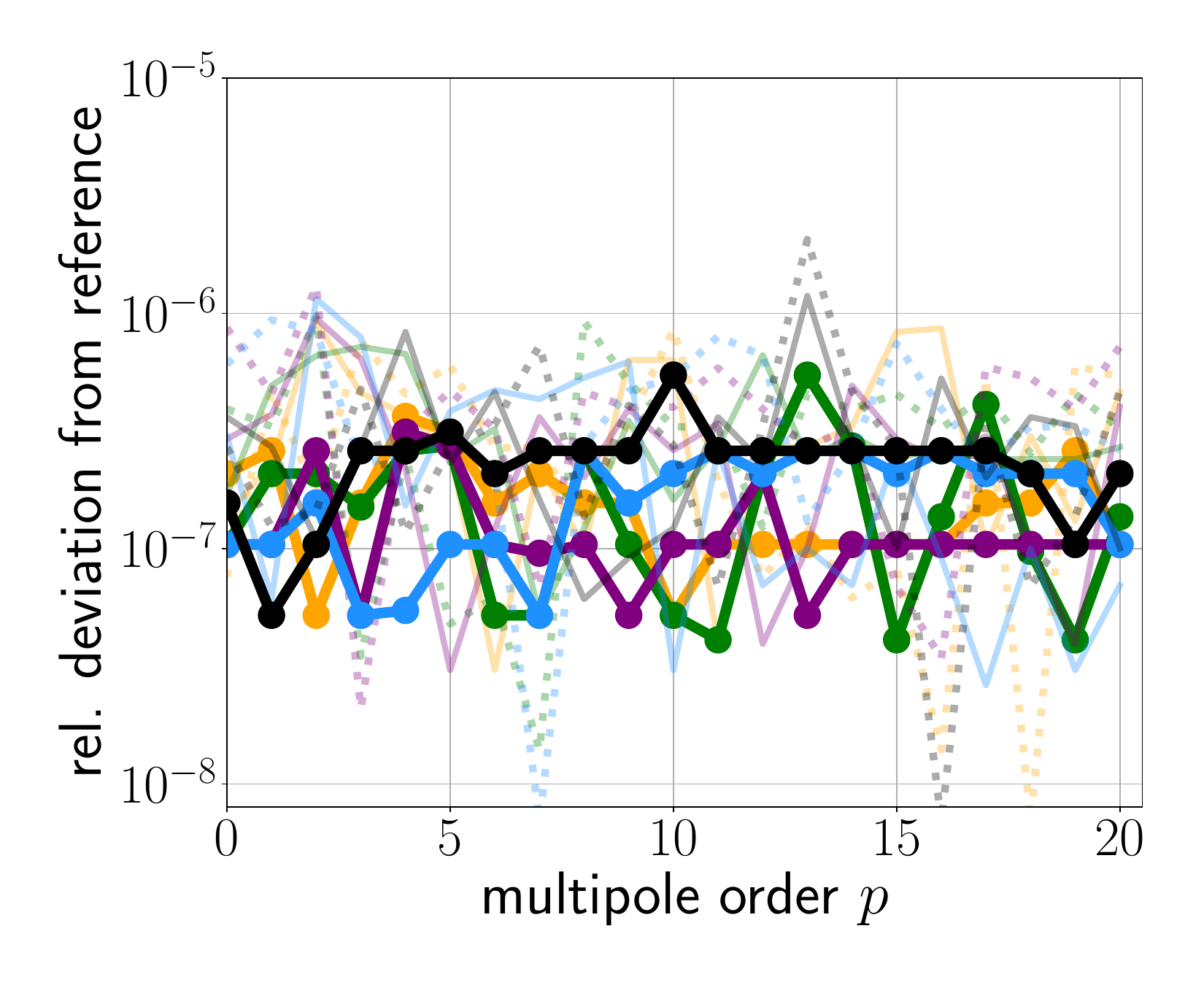}
        \caption{Single precision.}
        \label{fig:accuracy_single}
    \end{subfigure}

    \centering
    \begin{subfigure}[b]{0.49\textwidth}
        \centering
        \includegraphics[width=\textwidth]{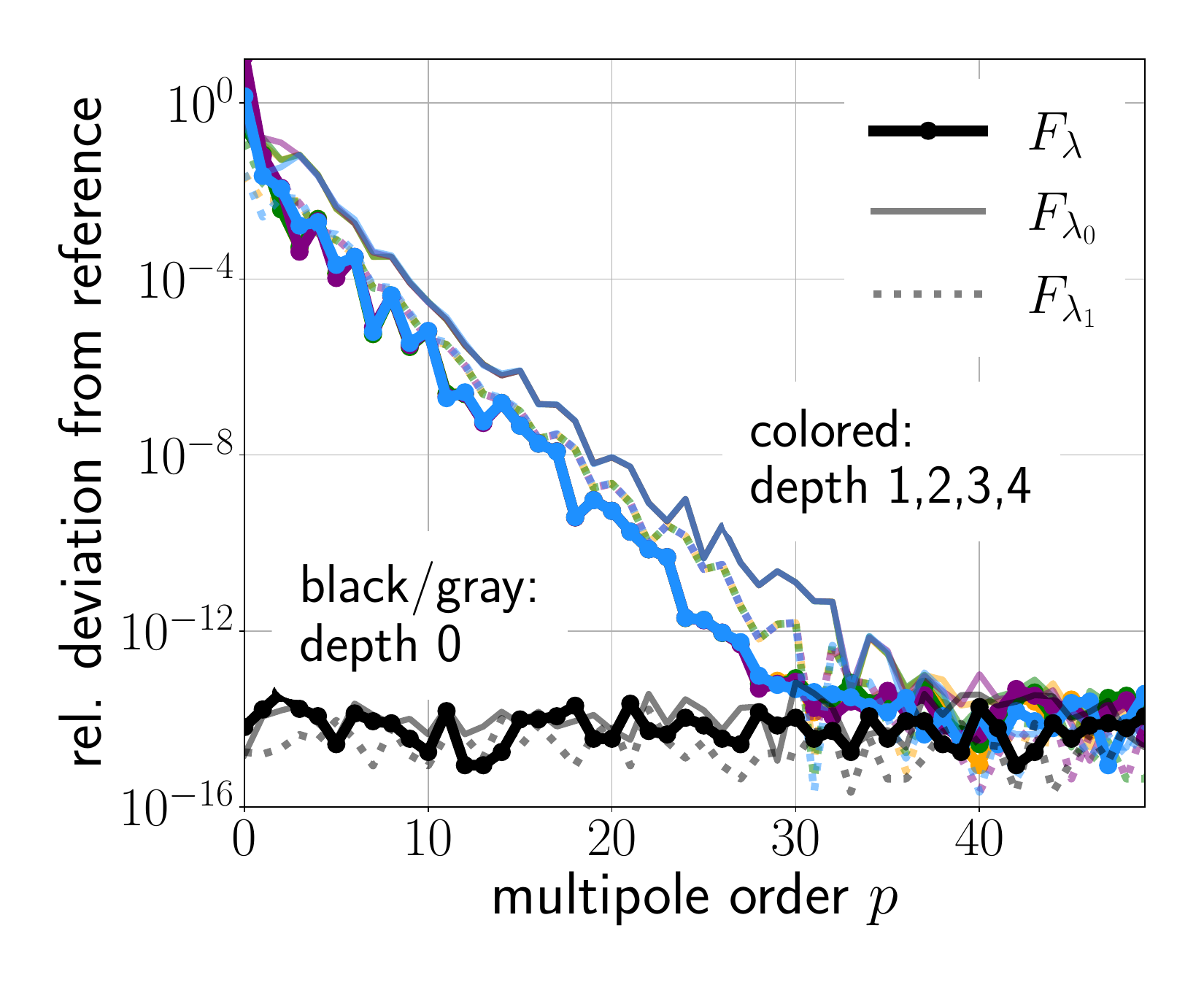}
        \caption{Double precision.}
        \label{fig:accuracy_double_worst_case}
    \end{subfigure}
    \begin{subfigure}[b]{0.49\textwidth}
        \centering
        \includegraphics[width=\textwidth]{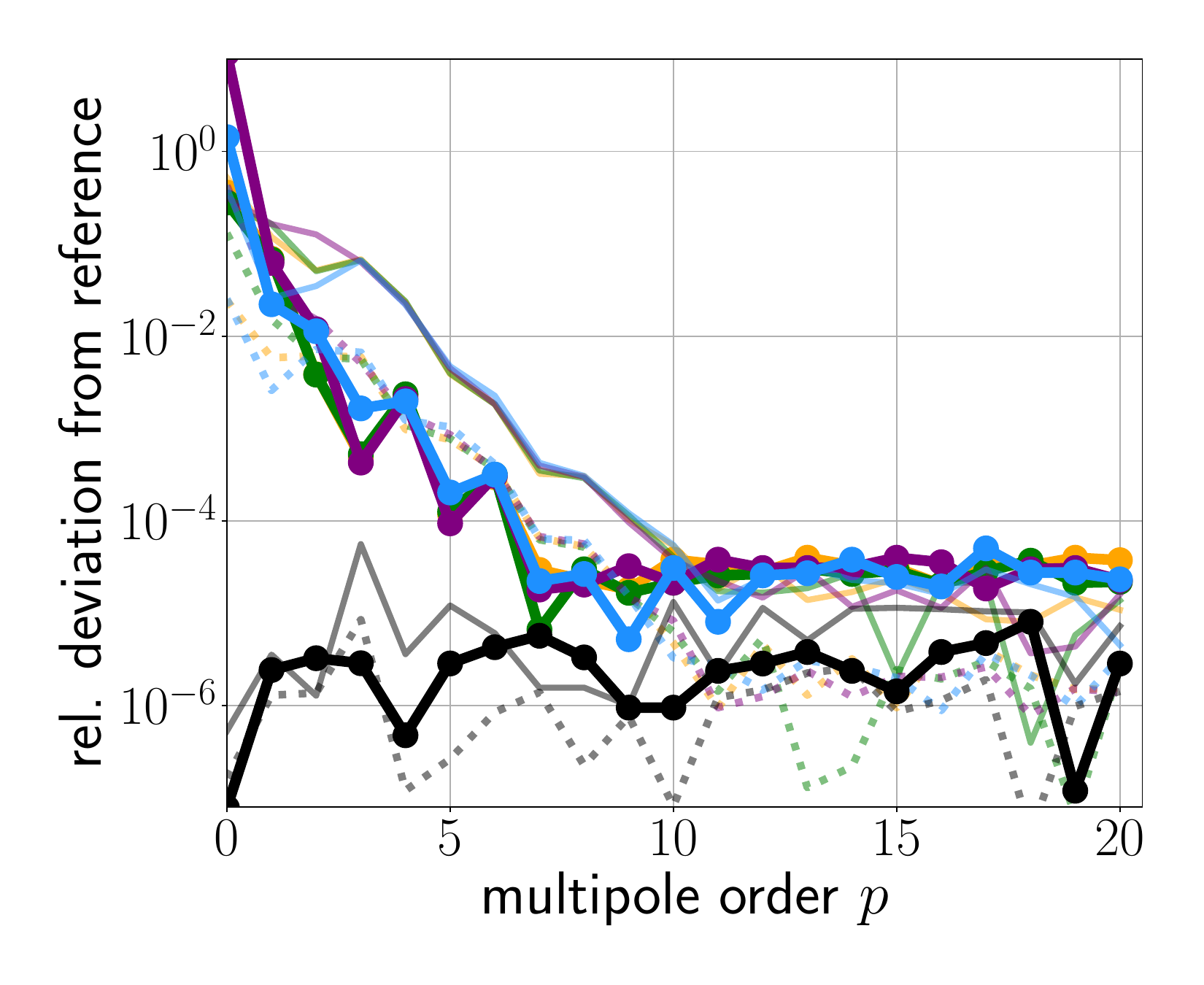}
        \caption{Single precision.}
        \label{fig:accuracy_single_worst_case}
    \end{subfigure}
    \caption{{\bfseries Quantification of the accuracy of the \fmmhi scheme.}
    The plots show the relative deviation of the \fmmhi forces from the reference forces
    for the 1,010 particle test system with typical particle distribution (a) and (b),
    and with hypothetical worst-case particle distribution (c) and (d).}
       \label{fig:accuracy}
\end{figure}

Figure~\ref{fig:accuracy} shows 
the relative deviations of forces $F_{\lambda}$ from the reference forces for two forms and $F_{\lambda_0}$, and $F_{\lambda_1}$ for four forms.
First the case of tree depth $d=0$ was considered.  
Here, all correction terms are evaluated directly,
hence \fmmhi is expected to be identical to the reference forces within numerical precision.
Indeed, the relative deviations for $d=0$ (depicted by the black and grey curves) are
within numerical precision for both double (a) and single precision (b) calculations for all considered multipole orders.

In contrast, for larger tree depths $d$, 
the colored curves in Figure~\ref{fig:accuracy} show differences to the reference forces particularly for low multipole orders $p$, 
which decrease as expected with increasing order.
The largest relative deviations (up to $10^{-6}$) are seen for $p\leq3$.
For $p \ge 28$, the relative deviation reaches numerical (double) precision,
indicating that the scheme does not introduce additional approximation errors. 

Interestingly, the maximum relative deviation for double precision is about $10^{-6}$, 
which is approximately single precision accuracy level.
Thus, one would expect no significant deviations and no dependence on multipole order when tested at single precision. 
In fact, as can be seen in Figure~\ref{fig:accuracy}b, the deviation remains at the same level over the entire range of $p$.

Next, the relative deviations of the forces on the $\lambda$ particle for the worst-case distribution of particles was quantified. 
Figure~\ref{fig:accuracy} shows a qualitatively similar behavior of the relative deviations in double precision compared to the more realistic case above. 
In particular, at tree depth $d=0$ the deviations remain at numerical (double) precision, 
whereas at larger depths increasing accuracy for increasing multipole order $p$ is achieved.
In contrast to the typical case, however, 
overall much larger deviations are seen, 
as expected for the worst-case particle distribution.

The observed convergence with increasing multipole order confirms that the deviations are due to 
the truncation of the multipole interactions. 
The larger deviations compared to the typical case can be attributed to the intended unfavorable non-clustered site particle distribution,
which implies that essentially none of the mutual interactions are calculated directly in the precalculated charge-scaled Hamiltonian.
As a result, all corrected interactions may have large deviations.
This is even more pronounced in single precision, 
where the relative deviations stay close to numerical (single) precision at tree depth $d=0$. 
However, for larger depths $d$, the relative deviations do not fall below $10^{-5}$.
We anticipate that this is due to large errors in multipole expansions built for only a few particles, 
and due to numerical cancellations that occur when summing of larger potential values with very small corrections.
 
Similar overall accuracies are seen for the whole $\lambda$ range (data not shown).

\subsubsection{Comparison of $\partial\mathcal{H} / \partial\lambda$ for FMM vs.\ PME}
\label{sec:deviation_from_PME}

\begin{figure}[h]
     \centering
     \begin{subfigure}[b]{0.49\textwidth}
         \centering
         \includegraphics[width=\textwidth]{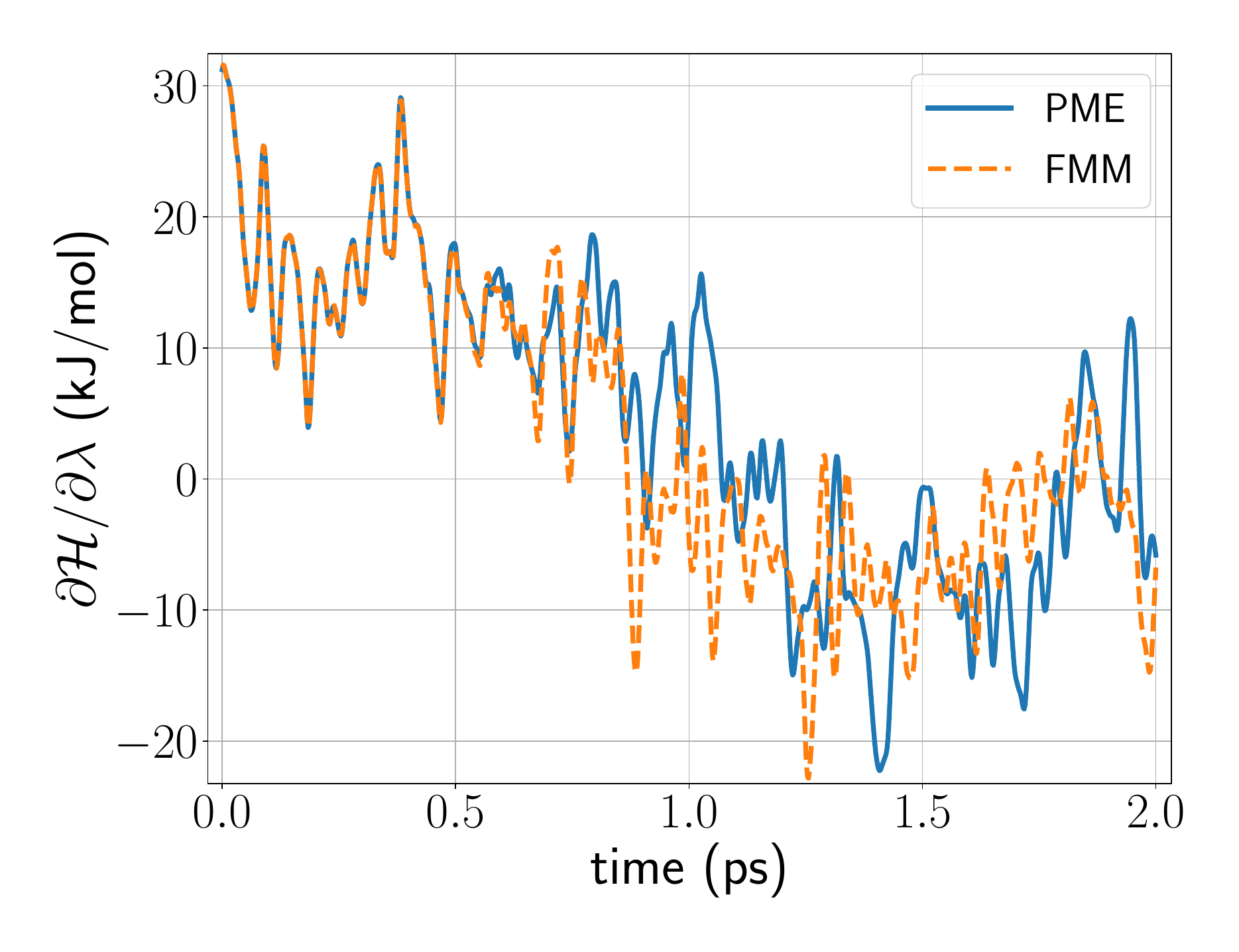}
         \caption{Separate trajectories.}
         \label{fig:benzene_test}
     \end{subfigure}
     \begin{subfigure}[b]{0.49\textwidth}
         \centering
         \includegraphics[width=\textwidth]{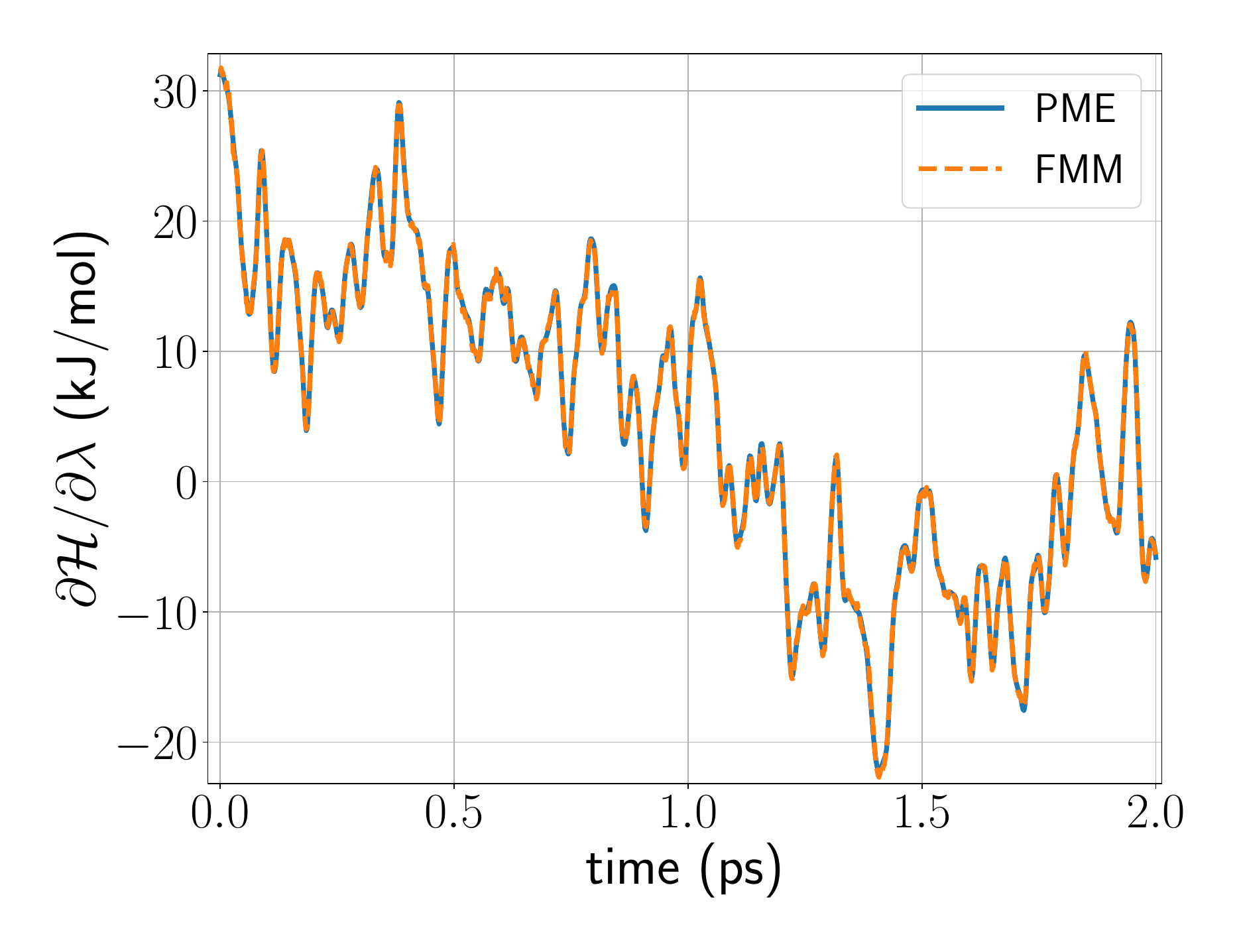}
         \caption{Precalculated atomic positions.}
         \label{fig:benzene_rerun_test}
     \end{subfigure}
     \caption{{\bfseries Comparison of the force acting on the $\lambda$ particle as calculated by PME and FMM
     for uncharging benzene.}
     The benzene ring carries its full charge at $t=0.0$~ps ($\lambda = 0$),
     while at $t=2.0$~ps ($\lambda = 1$) it is fully uncharged.
     FMM-computed $\partial\mathcal{H} / \partial\lambda$ values in orange, PME in blue.}
        \label{fig:benzene_accuracy}
\end{figure}

We next compared our FMM implementation to PME electrostatics using the benzene ring solvated in water. 
To this end, we used single precision FMM with multipole order $p=8$ and depth $d=3$.
Figure~\ref{fig:benzene_test} shows $\partial\mathcal{H} / \partial\lambda$
along a 2~ps trajectory calculated using PME (blue) and our FMM scheme (orange dashed), respectively. 
During this simulation, $\lambda$ covered the full range between zero and one.
As expected, essentially identical $\lambda$ trajectories are seen for the first several hundred integration steps; 
thereafter initially minute deviations amplify due to the chaotic nature of the dynamics of strongly interacting multiparticle systems. 
However, using exactly the same precomputed and stored input atomic positions for both PME and FMM, 
matching derivatives (Figure~\ref{fig:benzene_rerun_test}) were obtained over the entire $\lambda$ range. 
The root mean square error between FMM and PME forces is approximately $0.22$~kJ/mol.

\subsection{Performance}


Next, we assessed the computational performance of \fmmhi. 
Tho this end, we first characterized the scaling behavior with the number of sites, forms, and particles; 
second, we evaluated the overall performance
of \gromacs with corrected FMM is also evaluated in the context of constant pH \lambdadyn simulations.

\subsubsection{Scaling with Increasing Number of Sites}
\begin{figure}[H]
     \centering
     \begin{subfigure}[b]{0.49\textwidth}
         \centering
         \includegraphics[width=\textwidth]{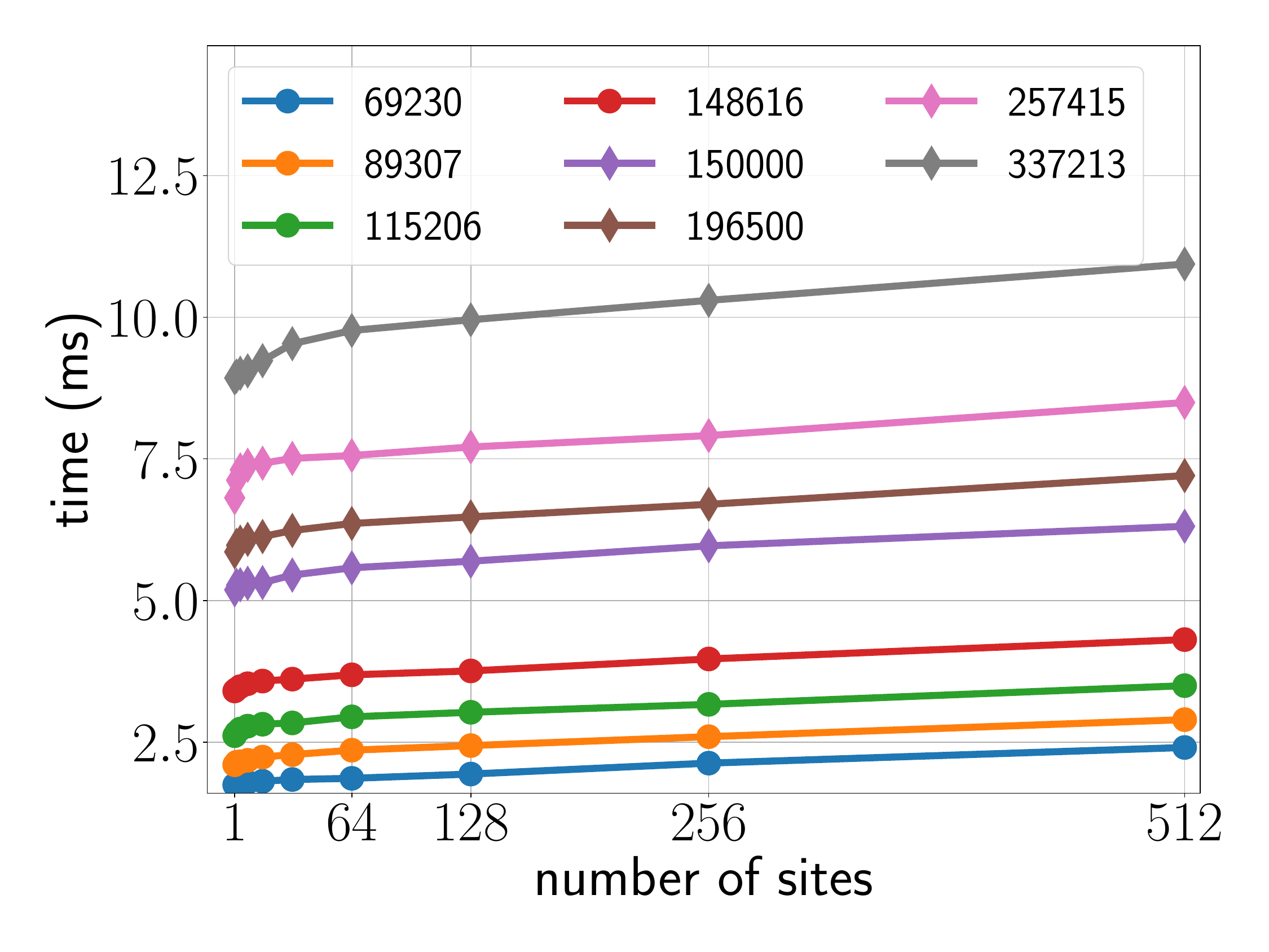}
         \caption{FMM runtime as a function of the number of sites for various particle counts as indicated in the top legend (diamonds for depth 3 and circles for depth 4).}
         \label{fig:sitesd3d4}
     \end{subfigure}
     \hfill
     \begin{subfigure}[b]{0.49\textwidth}
         \centering
         \includegraphics[width=\textwidth]{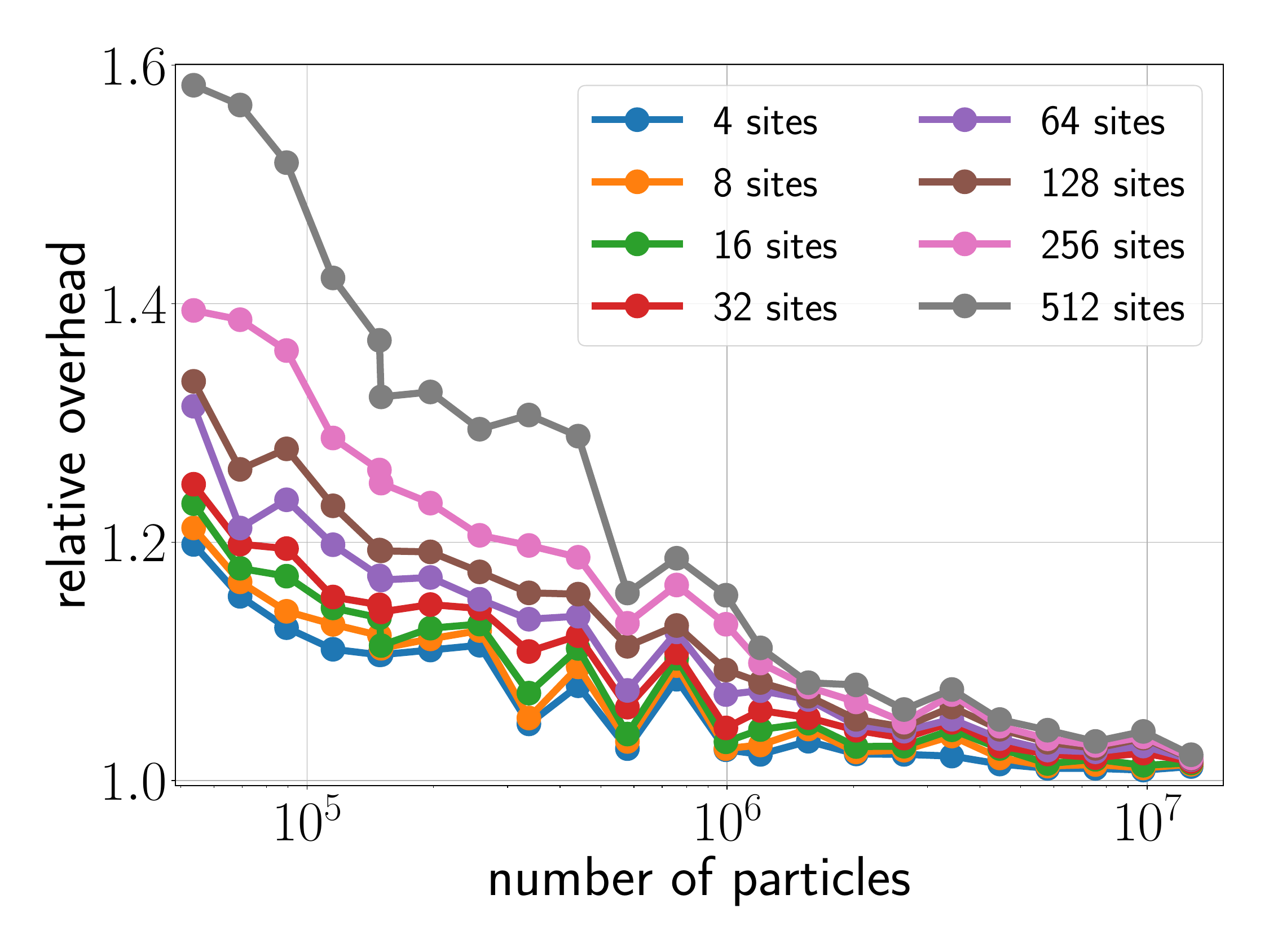}
         \caption{Relative overhead of adding new sites for different numbers of sites as a function of the number of particles. Baseline is a run without constant pH functionality.}
         \label{fig:sitesd3d4_relative}
     \end{subfigure}
     \caption{{\bfseries Scaling of the FMM based \fmmhi scheme for different numbers of sites and particles.}
     Results are shown only for the optimal tree depth.}
     \label{fig:sites_benchmark}
\end{figure}

As discussed in Subsection~\ref{sec:interaction_corrections}, 
we expect \fmmhi to scale linearly with the number of sites. 
Figure~\ref{fig:sitesd3d4} shows the average runtime for increasing the number of sites and for various particle counts between 70\,k to 340\,k at depths $d$ three and four, 
chosen for optimal performance.
For larger number of sites, 
the linear increase can clearly be seen, 
whereas for smaller number of sites a steeper increase is seen, 
due to constant costs associated with incorporating additional data structures and functions for correction calculations.
Moreover, for a small number of sites, 
constant pH related kernels do not achieve their optimal performance due to the insufficient computational load required to optimally utilize the underlying hardware. 
This effect can be seen more clearly in Figure~\ref{fig:sitesd3d4_relative},
which shows how the additional effort of the constant pH functionality scales with the number of particles for a given number of sites (different colors).
As can be seen, 
the relative performance overhead is quite small, 
and for realistic systems with moderate numbers of sites generally below 20\%. 
For larger systems of several million particles the overhead becomes negligible.


\subsubsection{Scaling with Number of Forms}
\begin{figure}[tb]
    \centering
         \includegraphics[width=0.6\textwidth]{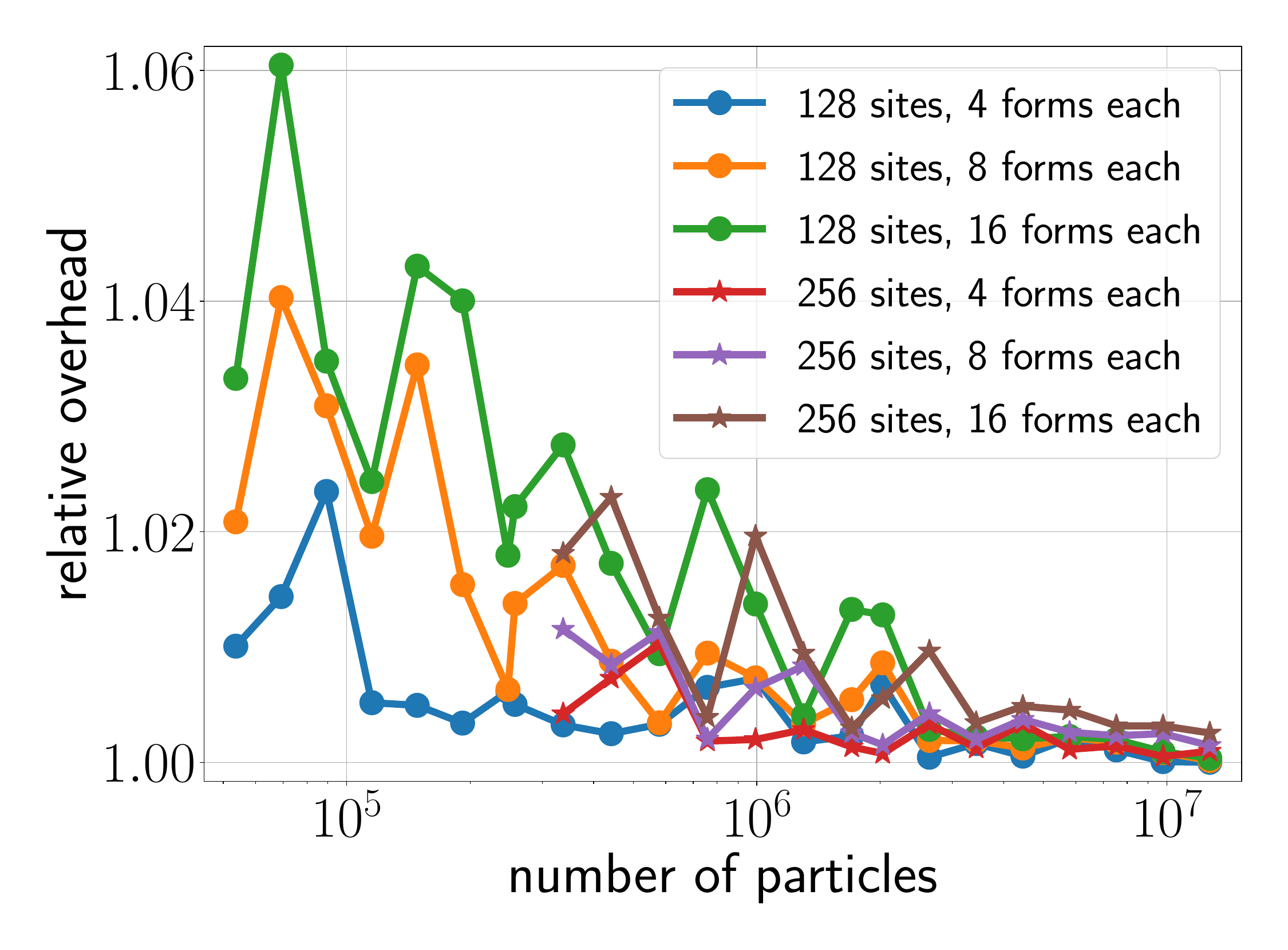}
    \caption{{\bfseries Costs of adding new forms to existing sites.}
    Results are shown for FMM tree depths of $d=3$ (circles) and $d=4$ (stars) for the random systems. 
    Baseline is the FMM performance with sites containing two forms (one $\lambda$ value).
    }
    \label{fig:forms_benchmark}
\end{figure}

Next, we studied how adding new forms to existing sites impacts performance.
Because additional forms require separate re-calculations (see Figure~\ref{fig:two_state_model}), 
one expects a moderate linear increase of the additional effort. 
Figure~\ref{fig:forms_benchmark} quantifies the resulting overhead for selected numbers of sites and forms (different colors).
Notably, adding new forms does not markedly affect performance, 
with an additional overhead of generally below 5\% and below 2\% for larger systems.
Here, the overhead is entirely due to the increase of calculations performed by the kernels. 
No additional data structures or kernel calls are required, which explains the small overhead.

\subsubsection{Scaling with Number of Particles}
\begin{figure}[tbp]
     \centering
     \begin{subfigure}[b]{0.65\textwidth}
         \centering
         \includegraphics[width=\textwidth]{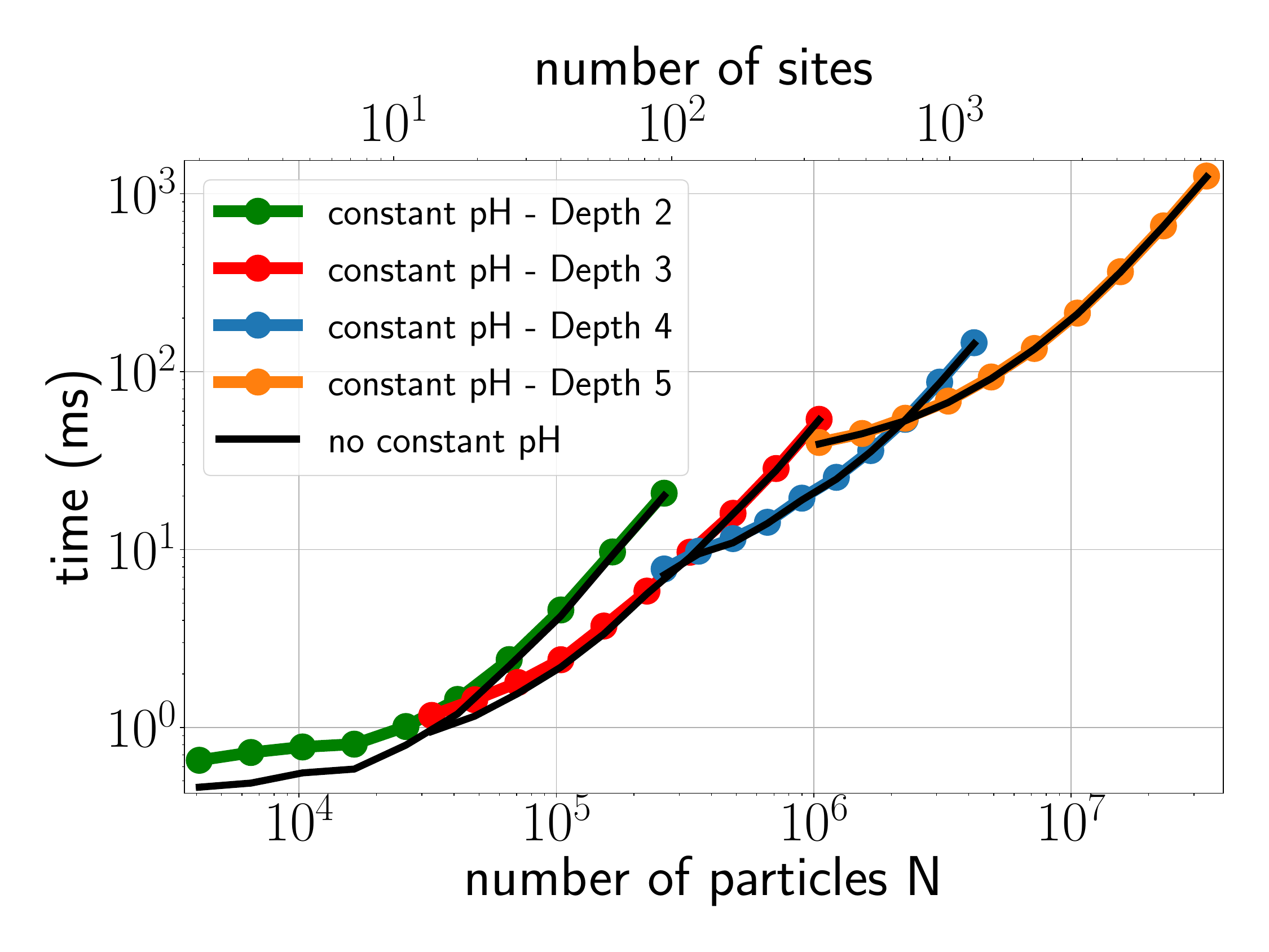}
         \caption{Absolute FMM runtime with (orange) and without (blue) constant pH.}
         \label{fig:constph_fmm_scaling}
     \end{subfigure}
     \begin{subfigure}[b]{0.65\textwidth}
         \centering
         \includegraphics[width=\textwidth]{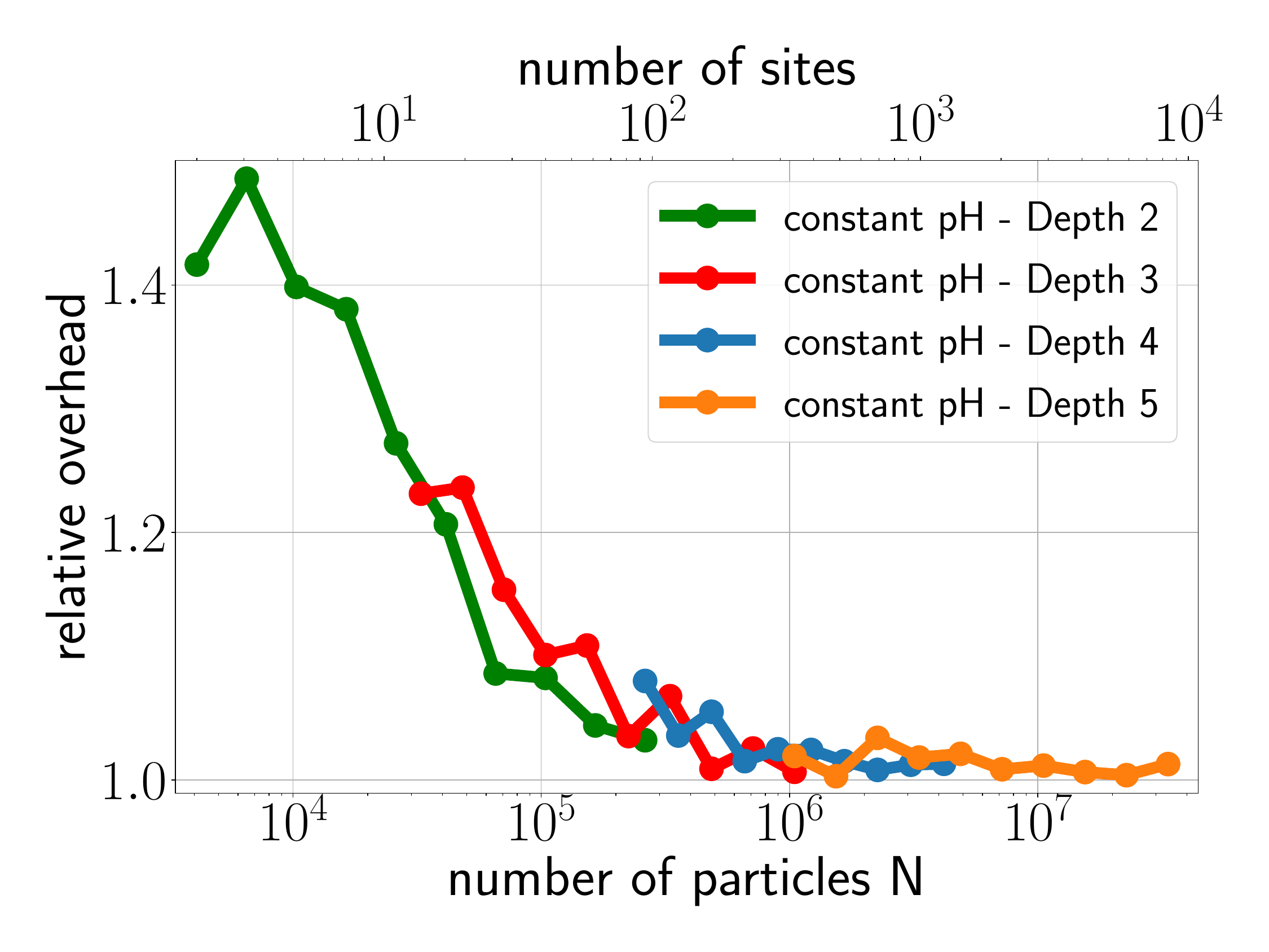}
         \caption{Relative overhead with constant pH versus a regular FMM.}
         \label{fig:constph_fmm_scaling_relative}
     \end{subfigure}
     \caption{{\bfseries Comparison of FMM runtime with and without constant pH.}
     This benchmark uses one site with ten particles for every 4,000 particles in the system, as estimated from lysozyme.}
     \label{fig:constph_fmm_scaling_benchmark}
\end{figure}

To characterize the scaling of the computational effort of \fmmhi with the number of particles under realistic simulation conditions,  
Figure~\ref{fig:constph_fmm_scaling} shows the absolute runtimes of FMM with and without constant pH functionality.
The characteristic behavior of FMM is evident in both cases,
with piecewise quadratic scaling for different choices of tree depth $d$. 
A proper choice of $d$ results in an overall linear scaling (dashed line) with system size.
Notably, the scaling of FMM is nearly unaffected by the constant pH overhead,
with small runtime differences seen only for systems with fewer than $10^5$ particles.
This finding is also reflected in the relative overhead of including constant pH (Figure~\ref{fig:constph_fmm_scaling_relative}), 
which decreases from approximately 50\% for very small systems (below $10^4$ particles) to below 10\% for typical system sizes,
and to nearly zero for large systems.
Overall, the addition of the constant pH feature has minimal impact on FMM scaling.
%
%

\subsubsection{\gromacs performance with FMM and constant pH}
\begin{figure}[h!]
    \centering
    \captionsetup[subfigure]{labelformat=empty}
    \begin{subfigure}[t]{0.499\textwidth}
        \centering
        \includegraphics[width=\textwidth]{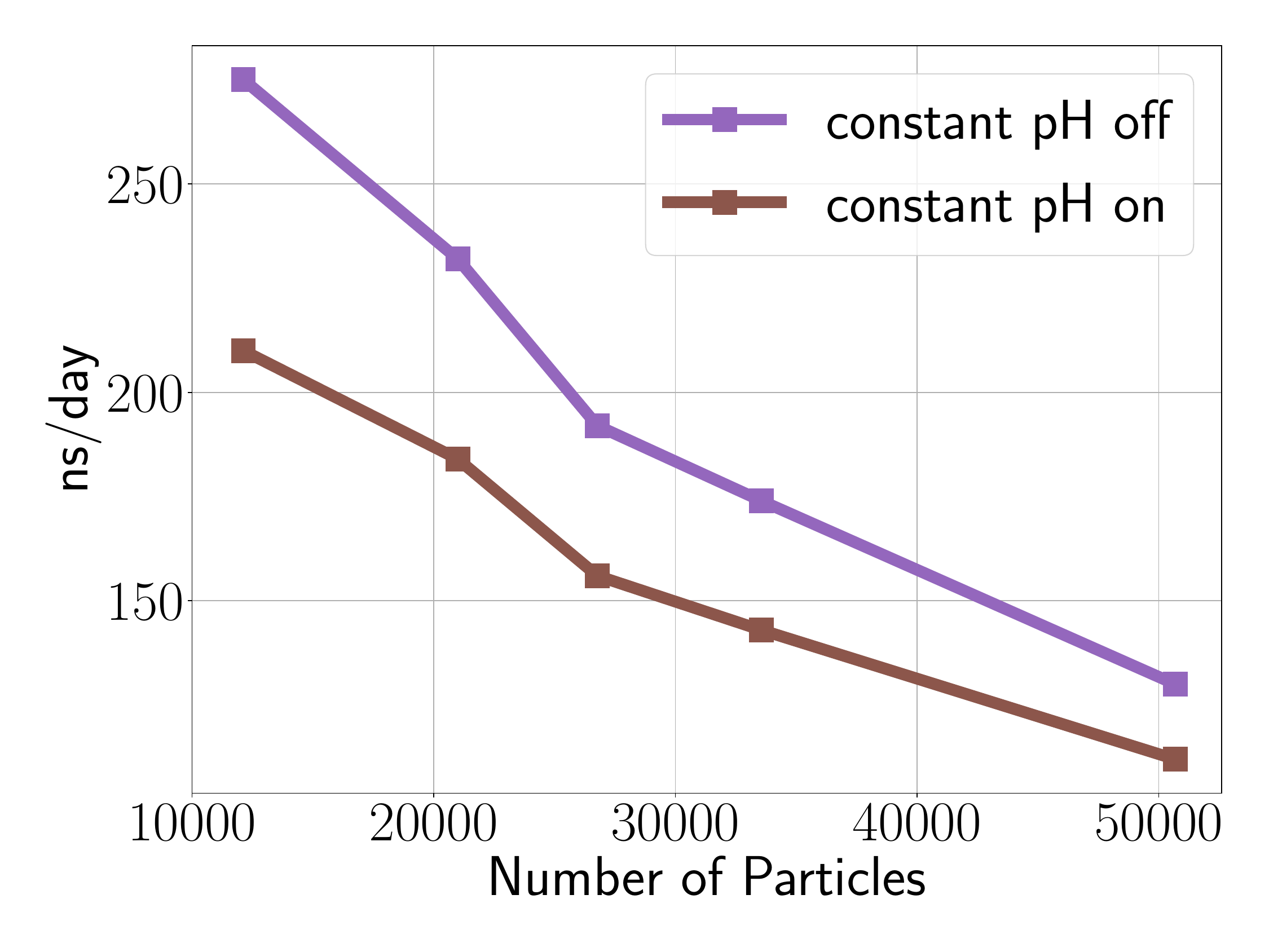}
        \caption{(a) Costs of \fmmhi-constant pH with FMM.}
    \end{subfigure}
     \hfill
    \begin{subfigure}[t]{0.49\textwidth}
        \centering
        \vspace{-6.2cm}
        \includegraphics[height=13.2cm]{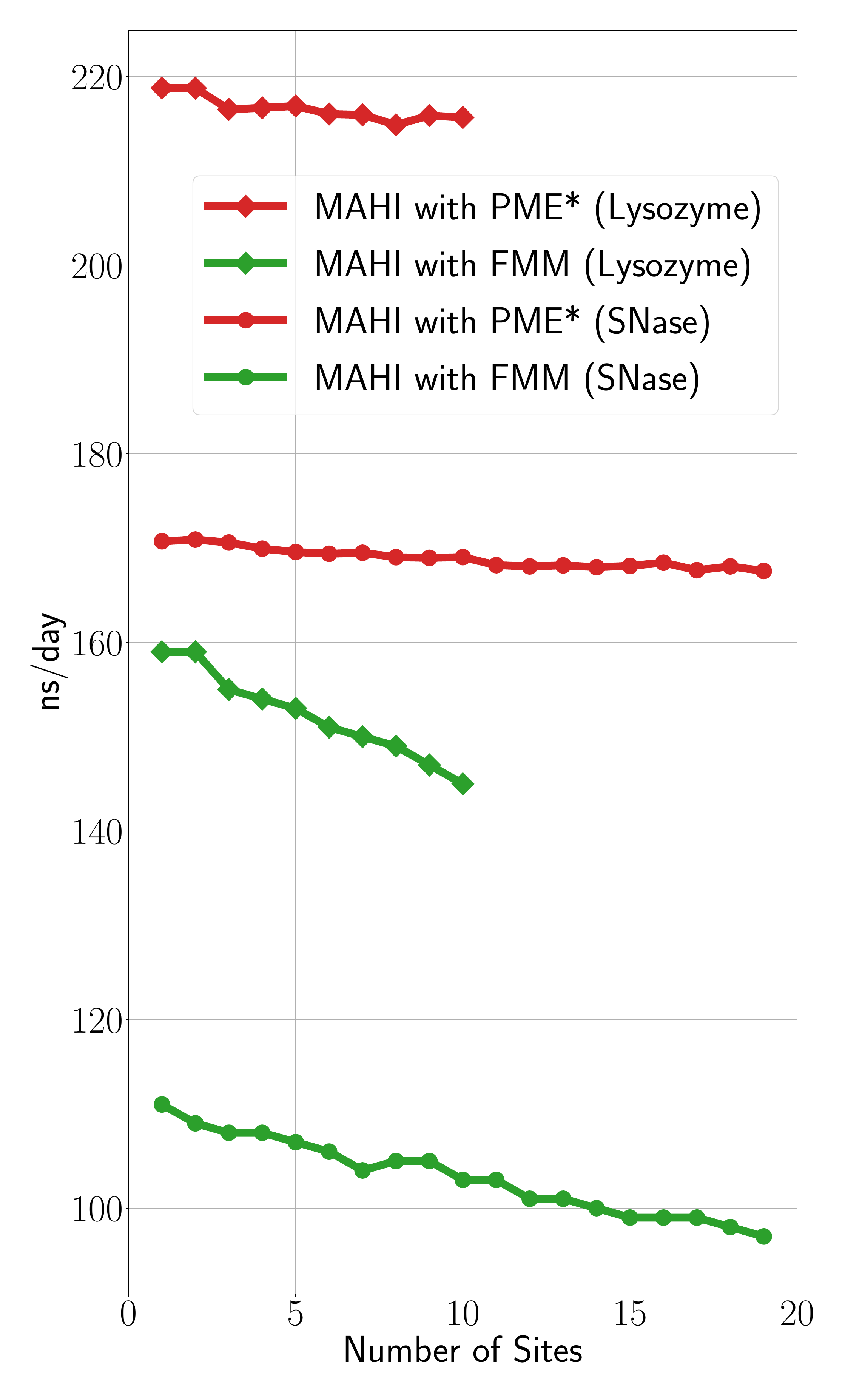}
        \vspace{0.1cm}
        \caption{(c) Absolute performances with \fmmhi.}
        \vspace{0.0cm}
    \end{subfigure}
    \begin{subfigure}[t]{0.499\textwidth}
        \centering
        \vspace{-6.9cm}
        \includegraphics[width=\textwidth]{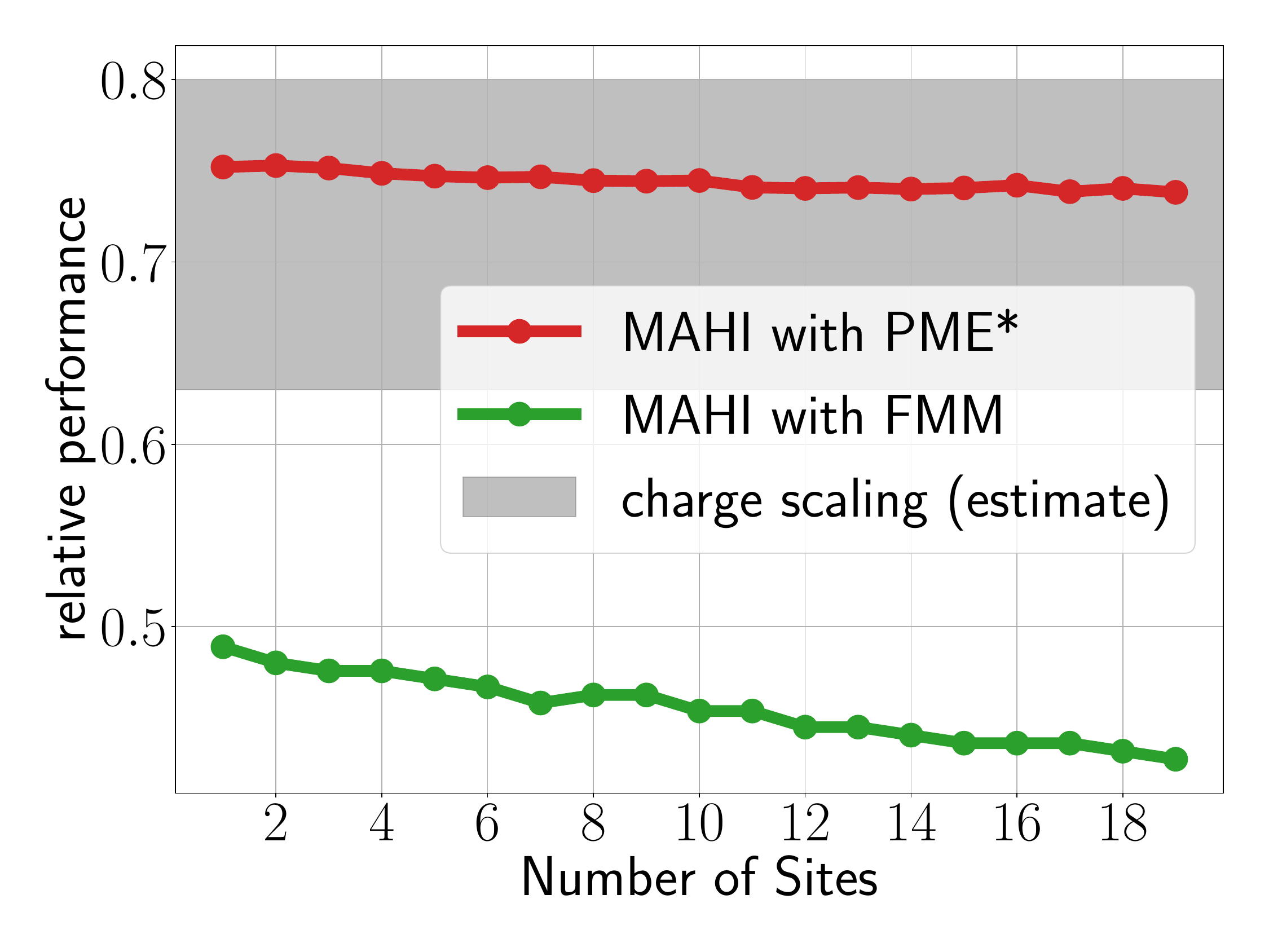}
         \caption{(b) Relative performance comparison.}
    \end{subfigure}
    \hspace{0.49\textwidth} 
    \caption{{\bfseries \gromacs performance with FMM and PME electrostatics for different constant pH simulation systems.}
    (a) A single titratable Glu in water water boxes of increasing size.
    (b) Relative performance for SNase with PME- and FMM-based MAHI compared to a fixed protonation simulation. 
    Costs of PME-based charge scaling estimated from Aho et.~al. (2022)\cite{aho2022constph} 
    (c) \gromacs performance with varying number of sites for FMM- and PME-based \fmmhi.
    *Preliminary results. 
    }
    \label{fig:constph_gromacs}
\end{figure}

The previous benchmarks assessed the performance of the constant pH FMM as a stand-alone solver.
Our final performance test, 
therefore, addressed the total runtime of a constant pH FMM \gromacs simulation, 
which also involves calculating the dynamics of the $\lambda$ particles.
This is likely the most relevant benchmark for most users.
Figure~\ref{fig:constph_gromacs} compares the constant pH performance to a regular \gromacs FMM run for
a single protonatable residue in a box of increasing size (a), and for lysozyme and SNase (c).
Here, too, a constant pH overhead of about 25\% is seen for small systems of about 10,000 particles, 
dropping below 10\% for systems with $\approx$~50,000 particles.
Additionally, 
the costs for adding new titratable sites range from 0.5 to 1 ns/day per site, 
as shown for small solvated proteins in Figure~\ref{fig:constph_gromacs}c.
Thus, for typical numbers of titratable sites in biomolecular systems the overhead due to the number of sites 
is negligible in both absolute and relative terms.
Overall, the computational effort of \fmmhi is essentially independent of the number of sites,
with only a minor impact (typically below 10\%) when increasing the number of sites.

Since the corrections $\Csr$ are applied to the charge-scaled potential $\mathcal{\tilde{V}}$ precalculated from a charge-interpolated system $\mathcal{\tilde{H}}$, 
it is actually irrelevant what method is used to obtain $\mathcal{\tilde{V}}$ as long as it is sufficiently accurate.
This finding opens up new routes for further performance improvements by using faster methods to obtain $\mathcal{\tilde{V}}$. 
Along this lines, we tested \fmmhi with PME such that HI-based $\lambda$ forces can be obtained.

Indeed, this hybrid approach proves advantageous in terms of overall performance. 
Figure~\ref{fig:constph_gromacs}c quantifies its performance, 
using the same test systems and parameters as above for the FMM-based \fmmhi.
We observed a 40\% and 55\% performance improvement for the lysozyme and SNase system, respectively.

%
%
%
%

Additionally, Figure~\ref{fig:constph_gromacs}b compares the performance of \fmmhi to the charge-scaling method in terms of relative overhead.
The performance of a charge scaling simulation (gray area in the panel) is in the range of 0.63--0.8 times the performance without constant pH, as estimated from Figure 6 A--B in \citet{aho2022constph}
PME-based \fmmhi (red curve) will also be in this range based on our performance estimates.
However, additional testing is required to confirm the accuracy of PME-based \fmmhi for constant pH simulations.

\subsubsection{Differences Between Hamiltonian and Charge Interpolation}
\label{sec:HIvsQI}

Having discussed the mathematical differences between HI and QI in Section~\ref{sec:HIvsQI_theory},
we will now assess the practical implications of these differences in constant pH simulations.

For a quantitative comparison of HI and QI, both constant pH setups must be equally well calibrated.
To this aim, 
the Glu reference compound in water at $\pH = \pKa$ was simulated for both HI and QI separately, 
using the acquisition protocol described in our companion publication.\cite{Briand2024}
In both simulations, 
the ratio of time spent in the protonated and deprotonated states, 
as well as the average $\lambda$ value, was $0.50 \pm 0.02$, allowing for a rigorous comparison.

\begin{figure}
\includegraphics[width=13.1cm]{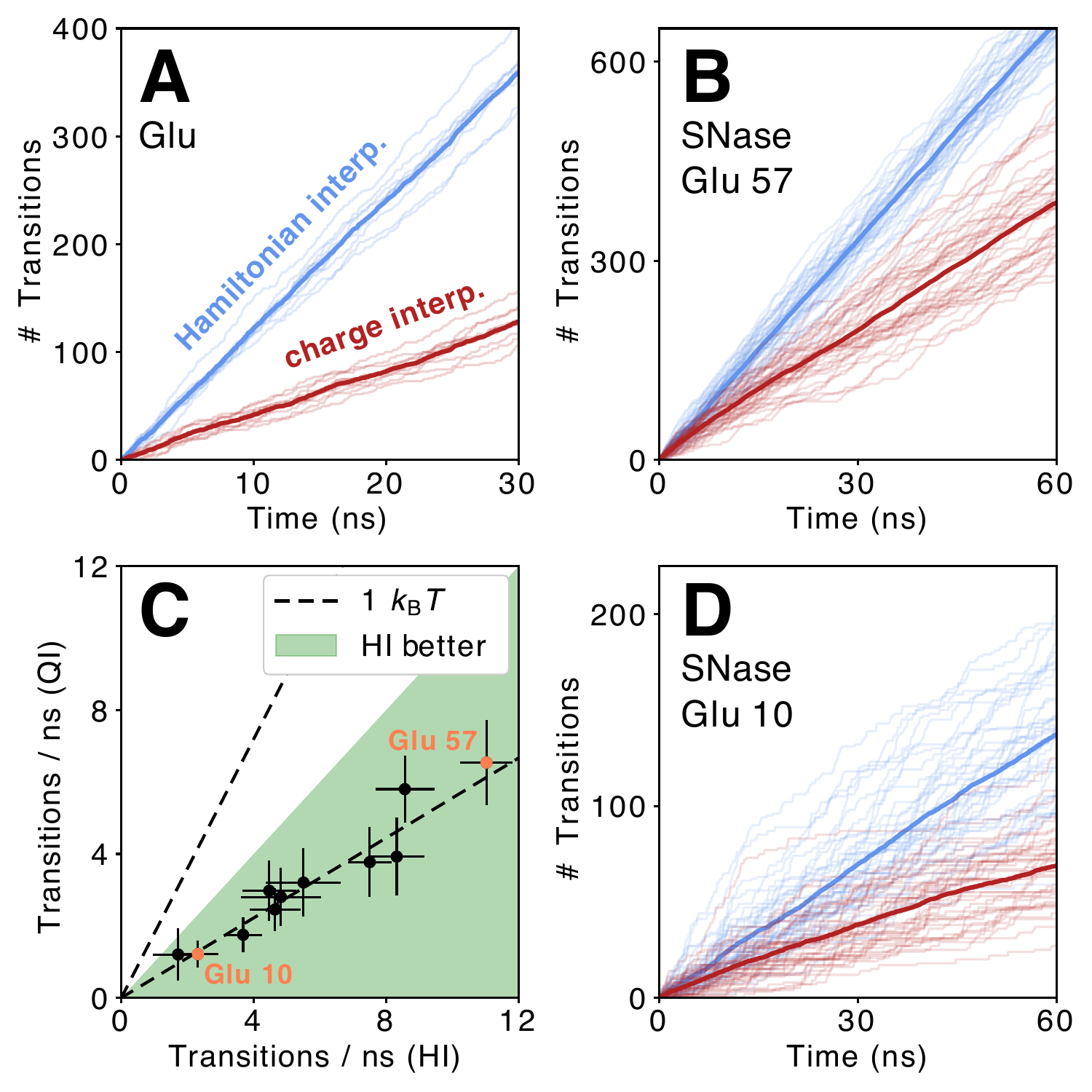} 
\caption{{\bfseries Comparison of Hamiltonian interpolation (HI) and charge interpolation (QI) constant pH simulations.}
Cumulative number of transitions for a single Glu residue in water (\textbf{A}),
and for Glu~57 (\textbf{B}) and Glu~10 (\textbf{D}) in the SNase protein.
Transparent lines correspond to individual replicas, solid lines to the average;
HI in blue and QI in red.
In (\textbf{C}), the transition rates of all Glu residues in SNase are compared.
Error bars give standard deviation across replicas.
Dashed line indicates how much an additional barrier of 1~$k_\text{B}T$ reduces the transition rate in a TST model.
}
\label{fig:QIHIGluRate}
\end{figure}

We started by investigating the differences for the single Glu residue in water.
Figure~\ref{fig:QIHIGluRate}A shows the cumulative number of transitions over time between protonated and deprotonated states for HI and QI.
With about twelve transitions per nanosecond for HI versus only four for QI, 
the transition rates are very different.
Using a Transition State Theory (TST) model,\cite{Eyring1935, laidler1983development}
this can be translated into an additional barrier of about 1~$k_\text{B}T$ for QI. 
This barrier corresponds to the harmonic potential identified in Section~\ref{sec:HIvsQI_theory}.

We then investigated the relevance of this additional barrier for larger proteins, using the SNase system. 
In the protein environment, transition rates vary for each residue due to the different local environments.
For instance, Glu~57 has the highest transition rate for both HI and QI (Figure~\ref{fig:QIHIGluRate}B and C), 
with a total of over 600 and 300 transitions, respectively, 
whereas Glu~10 shows less than 300 transitions in 60~ns (Figure~\ref{fig:QIHIGluRate}D).
While the transition rate for a given residue varies between replicas and over time, 
the average rate (Figure~\ref{fig:QIHIGluRate}C) is consistently higher for HI than for QI.
In the TST model, this difference corresponds to an additional barrier of about 1~$k_\text{B}T$
(black dashed line) as for the single Glu residue in water.

We therefore conclude that in practical simulations, in the absence of other variable factors
such as an automatic barrier optimization,\cite{Briand2024}
HI leads to higher transition rates than QI.
However, the magnitude of this effect, which is related to $k(\mathbf{r})$ (see eq~\ref{eqn:Udiff}), 
is a function of the parameterization of the residue of interest (charge, bond length) and can therefore vary.

\section{Conclusions}
Here, we derived and evaluated a constant pH \lambdadyn extension of our GPU-based FMM
that implements rigorous Hamiltonian Interpolation (HI).
It provides an efficient and scalable multipole-based computation of the difference between HI and precomputed charge-scaled Hamiltonians.
This implementation avoids redundant electrostatic calculations that typically arise in systems with many protonatable sites.
In addition, we demonstrate the integration of \fmmhi into the FMM framework and into the \gromacs software suite, 
enabling efficient and straightforward constant pH MD simulations.\cite{Briand2024}

We assessed the accuracy of the extension by comparing the forces acting on $\lambda$ particles to reference forces,
and found that these forces are within the accuracy range of the tested multipole order $p$, 
for both single and double precision.
In particular, the forces obtained are within numerical accuracy of those obtained by the \gromacs free energy module. 

Benchmarks of \fmmhi showed that for biomolecular applications, 
and particularly for moderate to large MD systems, 
the inclusion of typical numbers of titratable sites does not markedly affect the simulation performance, 
and that the involved computational overhead scales linearly with the number of sites and forms.
Benchmarks of the entire constant pH GROMACS implementation showed similar performance, 
demonstrating that pre- and post-handling of the data does not introduce any performance bottleneck. 
Overall, for a system comprising 100,000 particles, 
the overhead is less than 10\% compared to runs without constant pH.

To explore further ways to increase performance, we tentatively combined \fmmhi with PME.
In particular, we computed the charge-scaled Hamiltonian with PME, 
and then added the HI--QI difference with \fmmhi.
This exciting approach requires further testing and benchmarking, 
but promises another 40\% performance improvement. 
In addition to this practical benefit,  
this test demonstrates the flexibility of \fmmhi and shows that it can be combined with other electrostatic solvers such as PME,
which was previously deemed impractical.\cite{aho2022constph}
Initial tests indicate that PME-based MAHI simulations will not be much slower than charge scaling simulations.

Closer analysis of HI and QI revealed differences between the two interpolation schemes. 
We demonstrated that QI introduces protonation/deprotonation free energy barriers 
that are generally higher than those for HI. 
This reduces sampling efficiency and hinders the convergence of, e.g., \pKa calculations.
To mitigate this issue, we developed and evaluated an automatic barrier optimization protocol, 
described in our companion publication.\cite{Briand2024}

In addition to \fmmhi, 
FMM enables constant pH simulations of systems with open boundaries without further modifications. 
The combination of open boundaries and non-truncated treatment of long-range interactions is unique among fast electrostatics solvers currently used for MD.
FMM performs very well in this area, for example for droplet systems used in the simulation of mass spectrometry experiments.\cite{Marklund:2024}

As a next step, we aim to exploit the scaling properties of the FMM to enable constant pH for larger systems using multi-GPU, multi-node parallelism, in line with the trend towards exascale computing.
We anticipate that our constant pH approach will also scale favorably on exascale machines, 
as it is composed primarily of independent tasks that can be divided and parallelized with minimal communication overhead.
In particular, we expect that the parallelization of FMM avoids the considerable communication bottleneck and unfavorable scaling of the FFT required for PME.
Finally, \fmmhi is directly applicable not only to constant pH, 
but also more generally to all free energy computations that rely on the calculation of ${\partial\mathcal{H}}/{\partial\lambda}$.

A \gromacs version with the FMM-based constant pH module is available for download at
\url{https://www.mpinat.mpg.de/grubmueller/gromacs-fmm-constantph}.

\subsection*{Appendix}
In the following, we present the mathematical foundations that lead to the general formulation (eq.~\ref{eqn:iter_ham_lambda}) presented in this work.

As stated in eq~\ref{eqn:rek_ham_lambda}, the Hamiltonian
\begin{equation}
\mathcal{H} = (1-\lambda_{1})\left[(1-\lambda_{0})\mathcal{H}_{00} + \lambda_0\mathcal{H}_{01}\right] + \lambda_{1}\left[(1-\lambda_{0})\mathcal{H}_{10} + \lambda_0\mathcal{H}_{11}\right]
\nonumber 
\end{equation}
describes a system consisting of two sites with two forms each.
It is composed of four sub-Hamiltonians
$\mathcal{H}_{00}$, $\mathcal{H}_{01}$, $\mathcal{H}_{10}$, and $\mathcal{H}_{11}$,
\vspace*{-0.5cm}

\noindent \hspace*{-1.3cm}\makebox[\textwidth][l]{
\begin{minipage}{\textwidth}
\begin{align}
\mathcal{H}_{00} &= \color{black}\sum_{i=1}^{N^{(E)}} \sum_{j=i+1}^{N^{(E)}} \frac{{q}_i {q_j}}{r_{ij}} + 
\color{blue}\sum_{i=1}^{N^{(0)}}  \sum_{j=1}^{N^{(E)}} \frac{q^{0,0}_i q_j}{r_{ij}} + 
\color{blue}\sum_{i=1}^{N^{(1)}}  \sum_{j=1}^{N^{(E)}} \frac{q^{1,0}_i q_j}{r_{ij}} +
\color{green}\sum_{i=1}^{N^{(0)}}  \sum_{j=1}^{N^{(1)}} \frac{q^{0,0}_i q^{1,0}_j}{r_{ij}} + 
\color{red}\sum_{i=1}^{N^{(0)}}  \sum_{j=i+1}^{N^{(0)}} \frac{q^{0,0}_i q^{0,0}_j}{r_{ij}} +
\color{red}\sum_{i=1}^{N^{(1)}}  \sum_{j=i+1}^{N^{(1)}} \frac{q^{1,0}_i q^{1,0}_j}{r_{ij}}
\nonumber \\
\mathcal{H}_{01} &=\color{black}\sum_{i=1}^{N^{(E)}} \sum_{j=i+1}^{N^{(E)}} \frac{{q}_i {q_j}}{r_{ij}} +
\color{blue}\sum_{i=1}^{N^{(0)}}  \sum_{j=1}^{N^{(E)}} \frac{q^{0,1}_i q_j}{r_{ij}} + 
\color{blue}\sum_{i=1}^{N^{(1)}}  \sum_{j=1}^{N^{(E)}} \frac{q^{1,0}_i q_j}{r_{ij}} +
\color{green}\sum_{i=1}^{N^{(0)}}  \sum_{j=1}^{N^{(1)}} \frac{q^{0,1}_i q^{1,0}_j}{r_{ij}} + 
\color{red}\sum_{i=1}^{N^{(0)}}  \sum_{j=i+1}^{N^{(0)}} \frac{q^{0,1}_i q^{0,1}_j}{r_{ij}} +
\color{red}\sum_{i=1}^{N^{(1)}}  \sum_{j=i+1}^{N^{(1)}} \frac{q^{1,0}_i q^{1,0}_j}{r_{ij}}
\nonumber \\
\mathcal{H}_{10} &=\color{black}\sum_{i=1}^{N^{(E)}} \sum_{j=i+1}^{N^{(E)}} \frac{{q}_i {q_j}}{r_{ij}} +
\color{blue}\sum_{i=1}^{N^{(0)}}  \sum_{j=1}^{N^{(E)}} \frac{q^{0,0}_i q_j}{r_{ij}} + 
\color{blue}\sum_{i=1}^{N^{(1)}}  \sum_{j=1}^{N^{(E)}} \frac{q^{1,1}_i q_j}{r_{ij}} +
\color{green}\sum_{i=1}^{N^{(0)}}  \sum_{j=1}^{N^{(1)}} \frac{q^{0,0}_i q^{1,1}_j}{r_{ij}} + 
\color{red}\sum_{i=1}^{N^{(0)}}  \sum_{j=i+1}^{N^{(0)}} \frac{q^{0,0}_i q^{0,0}_j}{r_{ij}} +
\color{red}\sum_{i=1}^{N^{(1)}}  \sum_{j=i+1}^{N^{(1)}} \frac{q^{1,1}_i q^{1,1}_j}{r_{ij}}
\nonumber \\
\mathcal{H}_{11} &= %
\color{black} \underbrace{ \sum_{i=1}^{N^{(E)}} \sum_{j=i+1}^{N^{(E)}} \frac{{q}_i {q_j}}{r_{ij}}         }_{\text{env-env  }} +
\color{blue}  \underbrace{ \sum_{i=1}^{N^{(0)}} \sum_{j=1}^{N^{(E)}}   \frac{q^{0,1}_i q_j}{r_{ij}}                            +
\color{blue}               \sum_{i=1}^{N^{(1)}} \sum_{j=1}^{N^{(E)}}   \frac{q^{1,1}_i q_j}{r_{ij}}       }_{\text{env-site }} +
\color{green} \underbrace{ \sum_{i=1}^{N^{(0)}} \sum_{j=1}^{N^{(1)}}   \frac{q^{0,1}_i q^{1,1}_j}{r_{ij}} }_{\text{site-site}} +
\color{red}   \underbrace{ \sum_{i=1}^{N^{(0)}} \sum_{j=i+1}^{N^{(0)}} \frac{q^{0,1}_i q^{0,1}_j}{r_{ij}}                      +
\color{red}                \sum_{i=1}^{N^{(1)}} \sum_{j=i+1}^{N^{(1)}} \frac{q^{1,1}_i q^{1,1}_j}{r_{ij}} }_{\text{form-form}}  %
\color{black}.
\nonumber
\end{align}
\end{minipage}
}

\vspace*{0.2cm}
These sub-Hamiltonians describe the system in all four different protonation combinations of the two sites with two forms; 
both sites protonated (00), one of two sites protonated (01 and 10), and both sites deprotonated (11).

First, consider the interactions between the $\lambda$ independent \back particles (printed black in the above equations)
\begin{align}
\mathcal{H}_{\text{env-env}} &= (1-\lambda_{1})(1-\lambda_{0}) & \hspace{-8em} \sum_{i=1}^{N^{(E)}} \sum_{j=i+1}^{N^{(E)}} \frac{{q}_i {q_j}}{r_{ij}} \nonumber \\
                             &+ (1-\lambda_{1})\lambda_{0}     & \hspace{-8em} \sum_{i=1}^{N^{(E)}} \sum_{j=i+1}^{N^{(E)}} \frac{{q}_i {q_j}}{r_{ij}} \nonumber \\
                             &+ \lambda_{1}(1-\lambda_{0})     & \hspace{-8em} \sum_{i=1}^{N^{(E)}} \sum_{j=i+1}^{N^{(E)}} \frac{{q}_i {q_j}}{r_{ij}} \nonumber \\
                             &+ \lambda_{1}\lambda_{0}         & \hspace{-8em} \sum_{i=1}^{N^{(E)}} \sum_{j=i+1}^{N^{(E)}} \frac{{q}_i {q_j}}{r_{ij}}. \nonumber                              
\end{align}
Since
$(1-\lambda_{1})(1-\lambda_{0}) + (1-\lambda_{1})\lambda_{0} + \lambda_{1}(1-\lambda_{0}) + \lambda_{1}\lambda_{0} = 1$
and since the charges $q_i, q_j$ are independent of $\lambda$, the sums reduce to 
\begin{align}
\mathcal{H}_{\text{env-env}} = \sum_{i=1}^{N^{(E)}} \sum_{j=i+1}^{N^{(E)}} \frac{{q}_i {q_j}}{r_{ij}}. \nonumber
\end{align}
Next, consider the interactions between the site particles and \back particles (blue terms in the above equations)
\begin{align}
\mathcal{H}_{\text{env-site}} &=  (1-\lambda_{1})(1-\lambda_{0})& \hspace{-6em} \left(\sum_{i=1}^{N^{(0)}}  \sum_{j=1}^{N^{(E)}} \frac{q^{0,0}_i q_j}{r_{ij}} + 
\sum_{i=1}^{N^{(1)}}  \sum_{j=1}^{N^{(E)}} \frac{q^{1,0}_i q_j}{r_{ij}}\right) \hspace{1em} \nonumber \\
                              &+  (1-\lambda_{1})\lambda_{0}& \hspace{-6em} \left(\sum_{i=1}^{N^{(0)}}  \sum_{j=1}^{N^{(E)}} \frac{q^{0,1}_i q_j}{r_{ij}} + 
\sum_{i=1}^{N^{(1)}}  \sum_{j=1}^{N^{(E)}} \frac{q^{1,0}_i q_j}{r_{ij}}\right) \hspace{1em} \nonumber \\
                              &+  \lambda_{1}(1-\lambda_{0})& \hspace{-6em} \left(\sum_{i=1}^{N^{(0)}}  \sum_{j=1}^{N^{(E)}} \frac{q^{0,0}_i q_j}{r_{ij}} + 
\sum_{i=1}^{N^{(1)}}  \sum_{j=1}^{N^{(E)}} \frac{q^{1,1}_i q_j}{r_{ij}}\right) \hspace{1em} \nonumber \\
                              &+  \lambda_{1}\lambda_{0}& \hspace{-6em}     \left(\sum_{i=1}^{N^{(0)}}  \sum_{j=1}^{N^{(E)}} \frac{q^{0,1}_i q_j}{r_{ij}} + 
\sum_{i=1}^{N^{(1)}}  \sum_{j=1}^{N^{(E)}} \frac{q^{1,1}_i q_j}{r_{ij}}\right). \hspace{0.6em} \nonumber                           
\end{align}
For example, interactions between site 0 and \back are
\begin{align}
\bigg((1-\lambda_{1})(1-\lambda_{0}) + \lambda_{1}(1-\lambda_{0}) \bigg) &\sum_{i=1}^{N^{(0)}}  \sum_{j=1}^{N^{(E)}} \frac{q^{0,0}_i q_j}{r_{ij}} + \bigg((1-\lambda_{1})\lambda_{0} + \lambda_{1}\lambda_{0} \bigg)\sum_{i=1}^{N^{(0)}}  \sum_{j=1}^{N^{(E)}} \frac{q^{0,1}_i q_j}{r_{ij}}. \nonumber
\end{align}
Since $(1-\lambda_{1}) + \lambda_{1} = 1$ the higher $\lambda$ terms cancel out and the above expression reduces to 
\begin{align}
& (1-\lambda_{0}) \sum_{i=1}^{N^{(0)}}  \sum_{j=1}^{N^{(E)}} \frac{q^{0,0}_i q_j}{r_{ij}} + \lambda_{0} \sum_{i=1}^{N^{(0)}}  \sum_{j=1}^{N^{(E)}} \frac{q^{0,1}_i q_j}{r_{ij}}. \nonumber
\end{align}
Further, as the charges of the \back are independent of $\lambda_0$, the $\lambda_0$ terms can be put directly into the sums leading to charge-scaled interactions between site 0 and \back
\begin{align} 
\sum_{i=1}^{N^{(0)}} \sum_{j=1}^{N^{(E)}} \frac{((1-\lambda_{0})q^{0,0}_i + \lambda_0 q^{0,1}_i) q_j}{r_{ij}} \nonumber
= \sum_{i=1}^{N^{(0)}} \sum_{j=1}^{N^{(E)}} \frac{\tilde{q}^{(0)}_i q_j}{r_{ij}}. \nonumber                  
\end{align}

The same holds for site 1, 
hence all interactions between the sites and \back reduce to 
\begin{align}
\mathcal{H}_{\text{env-site}} &= \sum_{i=1}^{N^{(0)}} \sum_{j=1}^{N^{(E)}} \frac{\tilde{q}^0_i q_j}{r_{ij}} + \sum_{i=1}^{N^{(1)}} \sum_{j=1}^{N^{(E)}} \frac{\tilde{q}^{(1)}_i q_j}{r_{ij}}, \nonumber              
\end{align}
which are interactions between scaled charges and the \back.

Considering the interactions between particles of site 0 and of site 1 (site-site interactions, green color in the equations above), similarly to site-\back interactions, it holds
\begin{align}
\mathcal{H}_{\text{site-site}} &= \sum_{i=1}^{N^{(0)}} \sum_{j=1}^{N^{(1)}} \frac{\tilde{q}^{(0)}_i \tilde{q}^{(1)}_j}{r_{ij}}. \nonumber              
\end{align} 
This is valid for the same reason as in case of the site-\back interactions; for site 0 the charges $q^{0,0}$ and $q^{0,1}$ are $\lambda_{1}$ independent and for site 1 the charges $q^{1,0}$ and $q^{1,1 }$ are $\lambda_{0}$ independent.

When considering interactions between particles belonging to the same form of the same site (red terms in the above equations), all interacting charges depend $\lambda_0$ values.
Again, consider only site 0 for clarity.
\begin{align}
\mathcal{H}_{\text{form-form}} &=  (1-\lambda_1)(1-\lambda_0)& \hspace{-8em} \sum_{i=1}^{N^{(0)}}  \sum_{j=i+1}^{N^{(0)}} \frac{q^{0,0}_i q^{0,0}_j}{r_{ij}} \nonumber \\
                               &+  (1-\lambda_1)\lambda_0    & \hspace{-8em} \sum_{i=1}^{N^{(0)}}  \sum_{j=i+1}^{N^{(0)}} \frac{q^{0,1}_i q^{0,1}_j}{r_{ij}} \nonumber \\
                               &+  \lambda_1(1-\lambda_0)    & \hspace{-8em} \sum_{i=1}^{N^{(0)}}  \sum_{j=i+1}^{N^{(0)}} \frac{q^{0,0}_i q^{0,0}_j}{r_{ij}} \nonumber \\
                               &+  \lambda_1\lambda_0        & \hspace{-8em} \sum_{i=1}^{N^{(0)}}  \sum_{j=i+1}^{N^{(0)}} \frac{q^{0,1}_i q^{0,1}_j}{r_{ij}}. \nonumber                        
\end{align}
As in previous example, the $\lambda_1$ cancels out leading to 
\begin{align}
\mathcal{H}_{\text{form-form}} =  (1-\lambda_0) &\sum_{i=1}^{N^{(0)}}  \sum_{j=i+1}^{N^{(0)}} \frac{q^{0,0}_i q^{0,0}_j}{r_{ij}} +
                                \lambda_0\sum_{i=1}^{N^{(0)}}  \sum_{j=i+1}^{N^{(0)}} \frac{q^{0,1}_i q^{0,1}_j}{r_{ij}}. \nonumber             
\end{align}
Here, in contrast to the previous three parts, the entire interactions are scaled and the sums do not reduce to interactions between charge-scaled particles.

Now, consider a system with one site and four forms according to eq~\ref{eqn:rek_one_site_lambda}
\begin{equation}
\mathcal{H'} = (1-\lambda_{1})\left[(1-\lambda_{0})\mathcal{H'}_{00} + \lambda_0\mathcal{H'}_{01}\right] + \lambda_{1}\left[(1-\lambda_{0})\mathcal{H'}_{10} + \lambda_0\mathcal{H'}_{11}\right]. \nonumber
\end{equation}
Here, as in the previous examples, the partial Hamiltonians $\mathcal{H}_{\text{env-env}}$ and $\mathcal{H}_{\text{env-site}}$ reduce to interactions between scaled particles. 
The complete Hamiltonian reads
\begin{align}
\mathcal{H'} &=\sum_{i=1}^{N^{(E)}}\sum_{j=i+1}^{N^{(E)}} \frac{q_i q_j}{r_{ij}} + \sum_{i=1}^{N^{(0)}}\sum_{j=1}^{N^{(E)}} \frac{\tilde{q}^{(0)}_i q_j}{r_{ij}} \nonumber \\
			&+ \tilde{\lambda}^0 \sum_{i=1}^{N^{(0)}} \sum_{j=i+1}^{N^{(0)}} \frac{q^{0,0}_i q^{0,0}_j}{r_{ij}} \nonumber \\
            &+ \tilde{\lambda}^1 \sum_{i=1}^{N^{(0)}} \sum_{j=i+1}^{N^{(0)}} \frac{q^{0,1}_i q^{0,1}_j}{r_{ij}} \nonumber \\
            &+ \tilde{\lambda}^2 \sum_{i=1}^{N^{(0)}} \sum_{j=i+1}^{N^{(0)}} \frac{q^{0,2}_i q^{0,2}_j}{r_{ij}} \nonumber \\
            &+ \tilde{\lambda}^3 \sum_{i=1}^{N^{(0)}} \sum_{j=i+1}^{N^{(0)}} \frac{q^{0,3}_i q^{0,3}_j}{r_{ij}},  \nonumber                              
\end{align}
where only form-form interactions are weighted with the $\tilde{\lambda}$ values obtained as shown in eq~\ref{eqn:one_site_lambda}.

The differences in form-form interactions are the foundations for \fmmhi derivation, 
which is constructed as follows.  
For each site $\sigma = 1,\dots,M$ we construct a list 
\begin{equation}
\label{eqn:lambda_list}
\Omega^{\sigma}:= \left(\rule{0cm}{12px}(1-\lambda_{0}, \lambda_{0})_{0}, (1-\lambda_{1}, \lambda_{1})_{1},\dots,(1-\lambda_{L^{\sigma} - 1}, \lambda_{L^{\sigma} - 1})_{L^{\sigma} - 1}\right)
\end{equation} 
of length $L^{(\sigma)}:=\log_2(\#\Ss)$ containing pairs $(1-\lambda, \lambda)$ connecting distinct forms of the site $\Ss$, as in the multi-state model. 
Calculating the Cartesian product
\begin{equation}
\tilde{\Omega}^{(\sigma)}:= \left(\rule{0cm}{12px}\Omega^{(\sigma)}_0 \times \Omega^{(\sigma)}_{1} \times \dots\times \Omega^{(\sigma)}_{L^{}-1}\right) = \left(\rule{0cm}{12px}\tilde{\Omega}^{(\sigma)}_{\rho=0},\tilde{\Omega}^{(\sigma)}_{\rho=1},\dots,\tilde{\Omega}^{(\sigma)}_{\rho=|S^{(\sigma)}|-1}\right) 
\end{equation}
yields $\#\Ss$ lists of length $L^{(\sigma)}$.
Taking the products of all elements of each $\tilde{\Omega}^{(\sigma)}_{\rho}$
\begin{equation}
 \Lsr := \prod_{i=0}^{L^{(\sigma)}-1}\left(\tilde{\Omega}^{(\sigma)}_{\rho}\right)_i
\end{equation}
yields $\tilde{\lambda}$ values used in the general formulation. 
With this, the entire Hamiltonian, 
consisting of an arbitrary number of states $\mathcal{X}$ with an arbitrary number of forms per site $\mathcal{Y}$, 
can be transformed to into an equivalent general formulation 
where the $\lambda$'s are no longer factors in front of the sub-Hamiltonians, but weigh the different forms of a site.

\begin{figure}[tbp]
    \centering
    \includegraphics[width=0.6\textwidth]{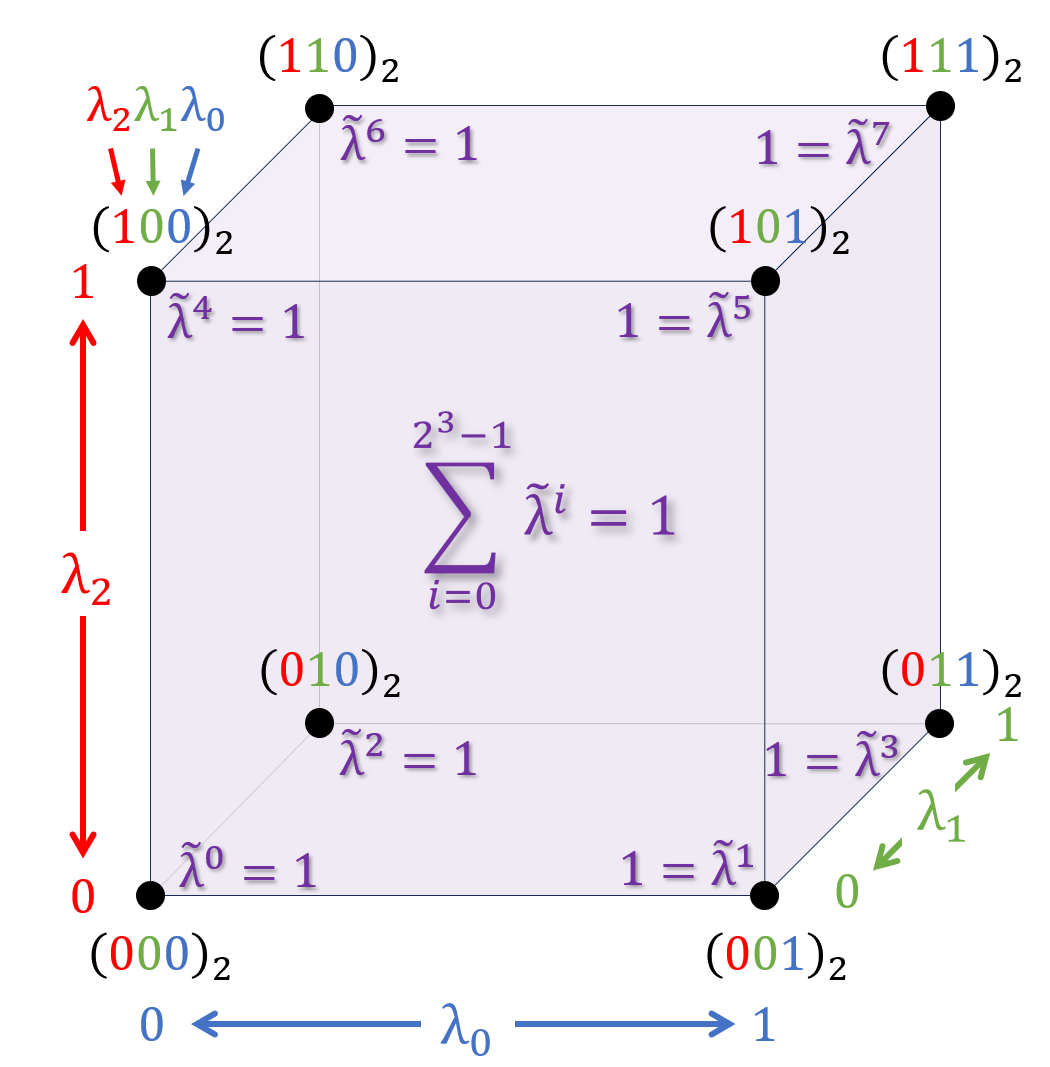}
    \caption{{\bfseries Sketch showing how $\lambda$ is mapped to $\tilde{\lambda}$.}
        Multi-state model for an exemplary site with eight states, demonstrating the mapping of $\lambda$ values to their corresponding $\tilde{\lambda}$ values.}
    \label{fig:two_state_model}
\end{figure}

The construction of multiple forms proceeds as follows.
Let us consider a system with one site (we omit the site index $\sigma$) and four forms 
\begin{equation}
\label{eqn:rek_one_site_lambda}
\mathcal{H'} = (1-\lambda_{1})\left[(1-\lambda_{0})\mathcal{H'}_{00} + \lambda_0\mathcal{H'}_{01}\right] + \lambda_{1}\left[(1-\lambda_{0})\mathcal{H'}_{10} + \lambda_0\mathcal{H'}_{11}\right].
\end{equation}
with prime notation to emphasize the difference to eq~\ref{eqn:rek_ham_lambda}.
Here, in contrast eq~\ref{eqn:rek_ham_lambda},
the sub-Hamiltonians $\mathcal{H'}_{\mathcal{Y}}$, where $\mathcal{Y}=\{00,01,10,11\}$, 
represent four different protonation forms of the same site.
Multiplying all $\lambda$ terms that belong to the same site yields
\begin{equation}
\label{eqn:one_site_lambda}
\mathcal{H}' = \underbrace{(1-\lambda_{1})(1-\lambda_{0})}_{:=\tilde{\lambda}^{(0)}}\mathcal{H'}_{00} + \underbrace{(1-\lambda_{1})\lambda_{0}}_{:=\tilde{\lambda}^{(1)}}\mathcal{H'}_{01} + \underbrace{\lambda_{1}(1-\lambda_{0})}_{:=\tilde{\lambda}^{(2)}}\mathcal{H'}_{10} + \underbrace{\lambda_{1}\lambda_{0}}_{:=\tilde{\lambda}^{(3)}}\mathcal{H'}_{11}.
\end{equation}
The index $\rho$ of each $\Lsr$ is a decimal representation of each binary element of ${\mathcal{Y}}$.
Figure~\ref{fig:two_state_model} depicts the enumeration of different site-forms and the corresponding $\lambda$ and $\tilde{\lambda}$ values for eight forms.

In practice, 
we encounter Hamiltonians $\mathcal{H}$ for multiple sites $\sigma =1,\dots,M$ that consist of sub-Hamiltonians $\mathcal{H}_\mathcal{X}$ and $\mathcal{H'}_{\mathcal{Y}}$. 
The straightforward multiplication of all $\lambda$ values (from different sites) to obtain $\Lsr$ for each site $\Ss$ according to eq~\ref{eqn:one_site_lambda} is still valid because the site-site interactions are calculated between charge-scaled particles, thus, the site independent $\lambda$ values cancel out.

To calculate the forces acting on the original $\lambda$ particles, the scaled correction terms $\Csr$ are subtracted from the corresponding energies derived from the interactions between particles of the same site-form $\Ssr$ only. 
To determine the derivatives $\partial\mathcal{H} / \partial\lambda^{(\sigma)}_i$, 
the indices $\rho$ of each site $\sigma$ are mapped back to the indices $i = 0,\dots,L^{(\sigma)}-1$. 
Table~\ref{tab:correction_mapping} shows the scheme for back-mapping eight corrections with indices $\rho = 0,\dots,7$ to three $\lambda$ values with indices $i = 0,1,2$.

This is achieved by continuously enumerating indices of $\lambda$-pairs created according to eq~\ref{eqn:lambda_list}
\begin{equation}
\mathcal{I}^{\sigma}:= \Bigl( (0,1)_{0}, (2,3)_{1}, \dots,(2(L-1), 2(L-1)+1)_{L-1}\Bigl), 
\end{equation}
and by taking the Cartesian product of these indices
\begin{equation}
\tilde{\mathcal{I}}^{(\sigma)}:= (\mathcal{I}^{(\sigma)}_0  \times \mathcal{I}^{(\sigma)}_{1} \times \dots\times \mathcal{I}^{(\sigma)}_{L-1}).
\end{equation}
This yields the index tuples of the original $\lambda$ values.
The calculation of the derivatives with respect to the original $\lambda$ values from the corrections $\Csr$
\begin{equation}
f(\tilde{\Omega}^{(\sigma)}_{\rho},k) = \prod_{\substack{l=0 \\ l \neq k}}^{L^{(\sigma)}-1} \left({\tilde{\Omega}}^{(\sigma)}_{\rho}\right)_l,
\end{equation}
excludes the contribution of the $\lambda$ value itself.
With this, we calculate intermediate correction terms (see Table~\ref{tab:correction_mapping})
\begin{equation}
\mathcal{K}^{\sigma}_{k} = \sum_{j=1}^{\Ns} \mathcal{\tilde{V}}_j q^{(\sigma,\rho)}_j - \sum_{\substack{\rho=0 \\ \tilde{\mathcal{I}}^{\sigma}_{\rho,k/2}=k}}^{|S^{(\sigma)}|-1} \Csr \;f(\tilde{\Omega}^{(\sigma)}_{\rho},k/2)  \;,\; k = 0,\dots,2L^{(\sigma)}-1,
\end{equation}
where $\mathcal{\tilde{V}}$ is the potential of the charge-scaled system defined in eq~\ref{eqn:iter_ch_ham_lambda}. 
Finally, the forces on the $\lambda$ particles are obtained with
\begin{equation}
\frac{\partial\mathcal{H}}{\partial\lambda^{(\sigma)}_i} = \mathcal{K}^{(\sigma)}_{\mathcal{I}^{(\sigma)}_{i,1}} - \mathcal{K}^{(\sigma)}_{\mathcal{I}^{(\sigma)}_{i,0}}, \quad i = 0,\dots,L^{(\sigma)}-1.
\end{equation}

\newcommand{\rcolBl}{\rowcolor{RoyalBlue!20}}
\newcommand{\rcolBr}{\rowcolor{BrickRed!20}}
\newcommand{\rcolY}{\rowcolor{Green!30}}
\begin{table}[tbp]
\caption{{\bfseries Mapping of the \fmmhi correction terms.}
Translation of the correction terms associated with the $\tilde{\lambda}$ values 
back to the terms for corrections of the forces on the initial $\lambda$ values,
here for an exemplary site with eight forms (enumerated 0 -- 7). 
The dots indicate which term applies to which original $\lambda$ force. }
\label{tab:correction_mapping}
\centering

\begin{tabular}{cccccccccccc}
\multirow{2}{*}{{\diagbox{$k$\hspace{1.4em}$i$}{\hspace{1.8em}$\rho$}}} & & & & 0  & 1 & 2 & 3 & 4 & 5 & 6 & 7          \\     
& & & & $(000)_2$ & $(001)_2$ & $(010)_2$ & $(011)_2$ & $(100)_2$ & $(101)_2$ & $(110)_2$ & $(111)_2$   \\ 
\arrayrulecolor{white}\hline\hline\hline
\vspace{-1.8cm}
\end{tabular}
\begin{tabular}{cccccccccccc}
\multirow{1}{*} & & & &        &        &        &       &        &        &         &           \\                                                  
 & & & & \phantom{$(000)_2$} & \phantom{$(001)_2$} & \phantom{$(010)_2$} & \phantom{$(011)_2$} & \phantom{$(100)_2$} & \phantom{$(101)_2$} & \phantom{$(110)_2$} & \phantom{$(111)_2$}   \\ 
\arrayrulecolor{white}\hline\hline\hline
\rcolBl 0 & 0 & $(1-\lambda_0)$ & $\leftarrow$ & $\bullet$ & $\bullet$ & $\bullet$ & $\bullet$ &           &           &           &             \\
\rcolBl 1 & 0 & $\lambda_0$     & $\leftarrow$ &           &           &           &           & $\bullet$ & $\bullet$ & $\bullet$ & $\bullet$   \\
\rcolBr 2 & 1 & $(1-\lambda_1)$ & $\leftarrow$ & $\bullet$ & $\bullet$ &           &           & $\bullet$ & $\bullet$ &           &             \\
\rcolBr 3 & 1 & $\lambda_1$     & $\leftarrow$ &           &           & $\bullet$ & $\bullet$ &           &           & $\bullet$ & $\bullet$   \\
\rcolY  4 & 2 & $(1-\lambda_2)$ & $\leftarrow$ & $\bullet$ &           & $\bullet$ &           & $\bullet$ &           & $\bullet$ &             \\
\rcolY  5 & 2 & $\lambda_2$     & $\leftarrow$ &           & $\bullet$ &           & $\bullet$ &           & $\bullet$ &           & $\bullet$   \\
\vspace{5mm}
\end{tabular}
\end{table}

\subsection*{Acknowledgments}
This work was financially supported by the German Federal Ministry of Education and Research (BMBF)
as part of the initiative "SCALEXA -- New Methods and Technologies for Exa\-scale Computing"
(BMBF project 16ME0713).
Thanks to Ivo Kabadshow for fruitful discussions on the FMM method and to Plamen Dobrev for his input on constant pH simulations.
The benzene MD system was kindly provided by Vytautas Gapsys.
We thank Gerrit Groenhof for his comments on the manuscript and his valuable suggestions.

\bibliography{biblio}

\end{document}